# Three Years of High-Contrast Imaging of the PDS 70 b and c Exoplanets at Hα with MagAO-X: Evidence of Strong Protoplanet Hα Variability and Circumplanetary Dust

Laird M. Close[1,2], Jared R. Males[2], Jialin Li[2], Sebastiaan Y. Haffert[2], Joseph D. Long[3], Alexander D. Hedglen[4], Alycia J. Weinberger[5], Kate Follette[6], Daniel Apai[2], Rene Doyon[7], Warren Foster[2], Victor Gasho[2], Kyle Van Gorkom[2], Olivier Guyon[2,8,9,10], Maggie Y. Kautz[9], Jay Kueny[9], Jennifer Lumbres[11], Avalon McLeod[12], Eden McEwen[9], Clarissa Pavao[13], Logan Pearce[2], Laura Perez[14], Lauren Schatz[15], J. Szulágyi[16], Kevin Wagner[2], Ya-Lin Wu[17]

## ABSTRACT

We present 3 years of high-contrast imaging of the PDS 70 b and c accreting protoplanets with the new extreme AO system MagAO-X as part of the MaxProtoPlanetS survey of Hα protoplanets. In 2023 and 2024 our sharp (25-27 mas FWHM); well AO corrected (20-26% Strehl), deep (2-3.6hr) images detect compact (r~30 mas; r~3 au) circumplanetary disks (CPDs) surrounding both protoplanets. Starlight scattering off the dusty outer edges of these CPDs is the likely source of the bright compact continuum light detected within ~30 mas of both planets in our simultaneously obtained continuum 668 nm filter images. After subtraction of contaminating continuum and PSF residuals with pyKLIP ADI and SDI we obtained high-contrast ASDI Hα images of both planets in 2022, 2023 and 2024. We find the Hα line flux of planet b fell by $(8.1\pm1.6)\times10^{-16}$ ergs/s/cm$^2$ a factor of 4.6 drop in flux from 2022 to 2023. In March 2024, planet b continued to be faint with just a slight 1.6x rise to an Hα line flux of $(3.64\pm0.87)\times10^{-16}$ ergs/s/cm$^2$. For c we measure a significant increase of $(2.74\pm0.51)\times10^{-16}$ ergs/s/cm$^2$ from 2023 to 2024 which is a factor of 2.3**x** increase. So both protoplanets have recently experienced significant Hα variability with ~1 yr sampling. In 2024, planet c is brighter than b: as c is brightening and b generally fading. We also tentatively detect one new point source "CC3" inside the inner disk (~49 mas; at PA~295°; 2024) with orbital motion roughly consistent with a ~5.6 au orbit.

---

[1] Corresponding author lclose@arizona.edu
Rest of the author affiliations at end of paper

# 1. INTRODUCTION

It is now well established that some gas giant protoplanets, pass through a period of high luminosity as they accrete hydrogen gas from their circumplanetary disks producing detectable Hα emission. This was most clearly demonstrated in the discovery of Hα emission from PDS 70 b (Wagner et al. 2018), and PDS 70 c (Haffert et al. 2019). Direct observations of protoplanets (defined here as accreting planets) are a key window into this poorly understood process of planet formation and accretion from a circumplanetary disk (embedded in a larger circumstellar disk). While the exact mechanisms of planetary accretion are not yet fully understood, massive planets could magnetospherically accrete, via magnetic fields, directly onto the polar regions of the planet (Zhu et al. 2016; Thanathibodee et al. 2019 and references within). Accretion through shocks onto the circumplanetary disk is also possible (Aoyama et al. 2018; Szulágyi & Mordasini 2017; Aoyama et al. 2021 and references within), and it is unclear which process, or a combination of both, dominate. Variability studies may be able to inform which of these models are more likely (Demars et al. 2023; and references within).

In section 2 of this manuscript we motivate the difficulty in directly detecting Hα emission from protoplanets. Then we briefly overview the upgraded Hα high-contrast mode (simultaneous/spectral differential imaging; SDI) of the MagAO-X instrument that was used in this work. In section 3, we describe the SDI observations of PDS 70 that were obtained in April 2022, March 2023 and March 2024. In section 4 we describe our new data reduction pipeline that was used to reduce these images to produce the final high-contrast images. In section 5, we analyze these images to produce the forward modeled photometry and astrometry for the protoplanets. Then in section 6, we carry out the analysis of the photometry to calculate the accretion luminosity, line flux at Hα, and mass accretion rates for the planets. In section 7 we have a general discussion about a select sample of interesting results from our dataset. First, we discuss the nature of the compact circumplanetary dust that is resolved around PDS 70 b and c. Then we discuss the variability of the PDS 70 protoplanets over 7 years (with MagAO, VLT/MUSE, HST, and this work with MagAO-X). We then explore if there are any outer planets beyond c. We then discuss if there are any inner planets inside of b's orbit. Finally, we discuss the nature of a "CC3" object detected at r~5.6 au. We present our conclusions in section 8.

## 2.0 MagAO-X INSTRUMENTAL CONFIGURATION FOR Hα IMAGING

### 2.1. Introduction to Why Imaging Hα Protoplanets is Technically Difficult

It is not trivial to detect protoplanets. The only way to guarantee an actively accreting protoplanet is being detected is to directly detect accretion tracers. Using the MagAO (the predecessor AO system to MagAO-X) system, Close et al. (2014) used the strongest visible tracer of accretion (Hα) to detect the low mass companion HD 142527 B inside the large transitional disk dust-free gap of HD 142527 A. They were able to confirm the presence of Hα by imaging in a narrowband (Δλ=6 nm) Hα filter and then subtracting the flux of the companion in a continuum narrowband (6 nm) filter (the MagAO SDI+ mode allowed both filters to be used simultaneously; Close et al. 2016). By combining angular differential imaging (ADI) and spectral differential imaging (SDI) Close et al. (2014) were able to obtain "ASDI" images of HD 142527B at just 80 mas from HD 142527 A. The excess of flux in the ASDI images proved that HD 142527 B indeed had significant Hα in emission and was accreting (see Balmer et al. (2022a, 2022b) for a recent detections and orbital solutions for HD 142527B).

Close et al. (2014) first speculated that for low mass (0.5<$M_{Jup}$<3) planets Hα ASDI imaging could be a powerful tool for detection of protoplanets, particularly at the lower mass end where Hα could be brighter than the NIR emission for active accretion. Indeed using MagAO's SDI+ mode Wagner et al. (2018) discovered Hα from the PDS 70 b protoplanet (Keppler et al. 2018) in May of 2018.

Another approach to detecting Hα protoplanets is to use an IFU. Indeed, Haffert et al. (2019) did use the VLT MUSE IFU and its laser guide star (LGS) AO system to detect PDS 70 b as well as discovered PDS 70 c with Hα emission lines that were distinct from that of the central star. A drawback to an IFU based approach is that the spaxels need to be large (~25 mas for MUSE) and so the spatial resolution is limited (~50 mas for MUSE). These large spaxels make it difficult to identify small dust structures in the circumplanetary environment.

Another successful approach to detecting Hα protoplanets is with *HST* where there is no need for AO correction. Indeed, Zhou et al. (2021) used the 2 nm narrowband Hα filter on *HST* to detect PDS 70 b. However, despite the ~100% Strehl of the *HST* PSF the small D=2.4m size of *HST* limited their angular resolution and inner working angle (IWA) such that PDS 70 c was not detected (Zhou et al. 2021). This left PDS 70c with only the detections of Haffert et al. at Hα. In a historical note PDS 70 c was also weakly detected at Hα in 2017 by MagAO by Follette et al. (2023) who re-reduced that older PDS 70 dataset.

Natural guide star (NGS) AO on the 8.2m VLT with SPHERE/ZIMPOL have yielded non-detections of PDS 70 b and c at Hα as well as no other new Hα planets (Cugno et al. 2019; Zurlo et al. 2020; Huélamo et al. 2022); however, HD 142527 B was re-detected. Hence, there has been a series of papers describing why Hα planets might be so rarely detected. Brittain et al. (2020) suggest that planetary accretion could be episodic in nature similar to an "FU Ori" type of outburst. Hence, it could be hard to catch the planets when they are near their peak of accretion/Hα luminosity. Brittain et al (2020) suggest that PDS 70 b and c are in the middle between the quiescent and burst state. If either planet dramatically increases (or decreases) its Hα brightness in the future this would strengthen the theory of Brittain et al. (2020). Hence, it is important to follow changes in the Hα line flux for these protoplanets. In another recent study, detailed 3D thermohydrodynamical simulations of Szulagyi & Ercolano (2020) show that the extinction from dust could extinguish Hα from all but the most massive ($\geq 10 M_{jup}$) planets. However, given that the masses of the PDS 70 c and b planets are ~2 and ~4 $M_{jup}$ (respectively; Wang et al. 2020), then the dust free "gas-only" models of Szulagyi & Ercolano (2020) are the only models in that study that can explain the observed properties and line strengths of PDS 70 b and c. In fact, a detailed physical model of magnetospherical accretion by Thanathibodee et al. (2019) shows that the accretion of PDS 70 b is well explained by magnetospherical accretion but the efficiency of Hα line luminosity productivity falls dramatically if the mass accretion rate falls below a certain crossing point (to be discussed more in section 6).

The PDS 70 protoplanets are located inside the large ~70 au dust free "disk gap" as imaged by ALMA. There is currently a tension in the literature as to how bright such "gap planets" should be at Hα. A key question that has not been rigorously poised or answered is: just how many of these Hα gap planets should we have detected already with current AO sensitivities? Are the null results (save PDS 70 b and PDS 70 c) significant --or simply a selection effect of the limits of the AO surveys themselves? Can we find any evidence that PDS 70 b and c are variable in brightness? Can we find any evidence of circumplanetary disks around b and c?

Unfortunately older AO systems (like MagAO, or SPHERE) cannot correct the atmosphere very well at Hα (656.3nm is quite blue for AO correction; Close et al. 2018), particularly with fainter guide stars, since the coherence patch size ($r_o$) of the atmosphere $r_o=22.5(\lambda/0.55)^{6/5}$ cm on a 0.5" seeing night. On such a night $r_o$ at H band is 84 cm but at H$\alpha$ it is just 28 cm. Therefore, only AO systems with ~14 cm sampling of the telescope primary mirror will Nyquist sample $r_o$ and make the highest contrast images at H$\alpha$ (see Close 2016 and Close et al. 2018 for reviews). For example, the Strehl of the corrected wavefront may be 75% (residual wavefront error 140nm rms) at H band ($\lambda$=1656 nm; where SPHERE was designed to work) but at H$\alpha$ it is <16% from just fitting error alone --so >84% of the starlight is outside of the diffraction PSF and is swamping any H$\alpha$ light from the planets. Moreover, for faint targets like PDS 70 the wavefront sensor and servo errors dramatically increase resulting in Strehls of ~1-4% for SPHERE at H$\alpha$ on PDS 70. This simple scaling has another "hit" for H$\alpha$ contrasts, the Strehls are so low that no coronagraph is used in any of the SHPERE/Zimpol datasets of Cugno et al. (2019) Zurlo et al. (2020) and Huélamo et al. 2022, hence the inner 0.2" of the SPHERE/ZIMPOL images have 100% of the diffracted and atmospheric speckles swamping the individual images (making contrasts of $\leq 10^{-4}$ at 100 mas impossible). Similar limits apply to MagAO's H$\alpha$ imaging as well (no coronagraph, low Strehls) and any NGS system. All these effects are not trivial and have made it difficult for MagAO to detect PDS 70 b even at 4$\sigma$ and impossible for SPHERE/ZIMPOL for PDS 70b (or c) at H$\alpha$ (Wagner et al. 2018). The 4$\sigma$ detection of PDS 70 b by MagAO was confirmed by VLT/MUSE by Haffert et al. (2019) in excellent ~0.4" seeing utilizing its powerful laser guide star.

In all these cases, the detections were difficult and required excellent atmospheric conditions, and good AO correction on PDS 70 which is not a very bright natural guide (NGS) star at R~11.7 mag (I~10.5 mag is actually more meaningful as I band ($\lambda$=806nm) is closer to the wavelengths (750 nm<$\lambda$< 950 nm) where the MagAO-X WFS is actually working; see Fig. 1). Hence, it is fair to ask: Is it a selection effect that, to date, most of the failed searches for H$\alpha$ planets is due to sub-optimal instrumentation for H$\alpha$ high contrast imaging? In this manuscript we seek to answer the question: exactly how well one can detect H$\alpha$ planets with modest Strehls (~20%) if one also utilizes a highly optimized H$\alpha$ high-contrast instrument on a large telescope (D=6.5m), leveraging dual photon-counting cameras, and an optimized KLIP ASDI reduction pipeline? This will be the approach of our MaxProtoPlanetS survey described briefly below.

## 2.2. New Hα detection Techniques with Extreme Visible AO: MagAO-X

Past "Hα AO" detections were done with older AO systems (VLT/SPHERE, VLT/MUSE, Magellan/MagAO) with relatively low (<1-10%) Strehls at Hα. However, we have now fully commissioned the world's newest extreme AO system MagAO-X. MagAO-X is unique --it was designed from the start to work in the visible at high Strehl (Males et al. 2018; 2024). The optical design for MagAO-X is complex in that being a woofer-tweeter system requires 2 reimaged pupils and then the lower coronagraphic bench (see Fig. 1) requires another pre-apodizer pupil followed by a Lyot pupil plane, hence MagAO-X has 4 reimaged pupils created by 8 off-axis parabolas (OAPs). The OAP relays are designed to minimize wavefront aberration on and off axis (out to a 6x6" FOV). We achieved this by successively slowing the f/ratio (and OAP off-axis angles) down as light moves through the instrument (from the Magellan f/11.04 → f/16 → f/57 → f/69) using extremely well-polished protected silver custom OAPs, (all flats were super-polished $\lambda/50$ surface, or better, optics as measured post protective silver coating) in this manner we simultaneously minimize both alignment errors in focal, and pupil, planes. The FOV is kept completely diffraction-limited (Strehl>95% over a 1.1" FOV) by roughly matching each OAP pair's f/ratios to the inverse of the OAP angle ratios and clocking each OAP so off-axis aberrations cancel in each relay, this is critical to eliminate field aberrations that cannot be corrected by AO. Since MagAO-X is not isothermal we minimized temperature related "misalignment creep" by having all the mirrors and beamsplitters housed in our custom patented (patent US11846828B2) micro-radian stable (<0.5 μrad/C), all stainless steel, locking kinematic mounts.

In summary, we have achieved an undistorted (and ghost free) 6x6" FOV at f/69 with 0.00590"/pix platescale (with 13 micron EMCCD pixels) that yields a nicely oversampled 3.4 pix/($\lambda$/D) at Hα. See Close et al. (2018) for more detail about the optical design of MagAO-X.

MagAO-X yields a superior level of wavefront control with a 2040 actuator Tweeter deformable mirror (DM) and a unique "extra" DM to eliminate all Non-Common Path (NCP) errors between the science and wavefront sensing channels, minimizing coronagraphic leak (we call this DM the NCP DM). This NCP DM was 97 elements, but in 2024 was upgraded to 1024 actuators which greatly improved our ability to use Focal Diversity Phase Retrieval (FDPR; Van Gorkom et al. 2021; Kuney et al. 2024) . Wavefront sensing (WFS) with MagAO-X's very low noise (<0.6 rms e- read noise) EMCCD pyramid WFS OCAM2 detector allows Strehls of >50% to be obtained at z' (910 nm; Δλ=130 nm) while closed loop at 2kHz (residual WFE <120nm rms with 1564 corrected modes– as demonstrated on-sky (Males et al. 2022). The low noise of this sensor allows good correction even on faint I~11 mag guide stars in good 0.5" seeing conditions. The MagAO-X system with up to 1564 corrected modes maps to ~14 cm/actuator, making it the highest sampled AO system in the world. So deeper, much more sensitive surveys for Hα planets are finally possible.

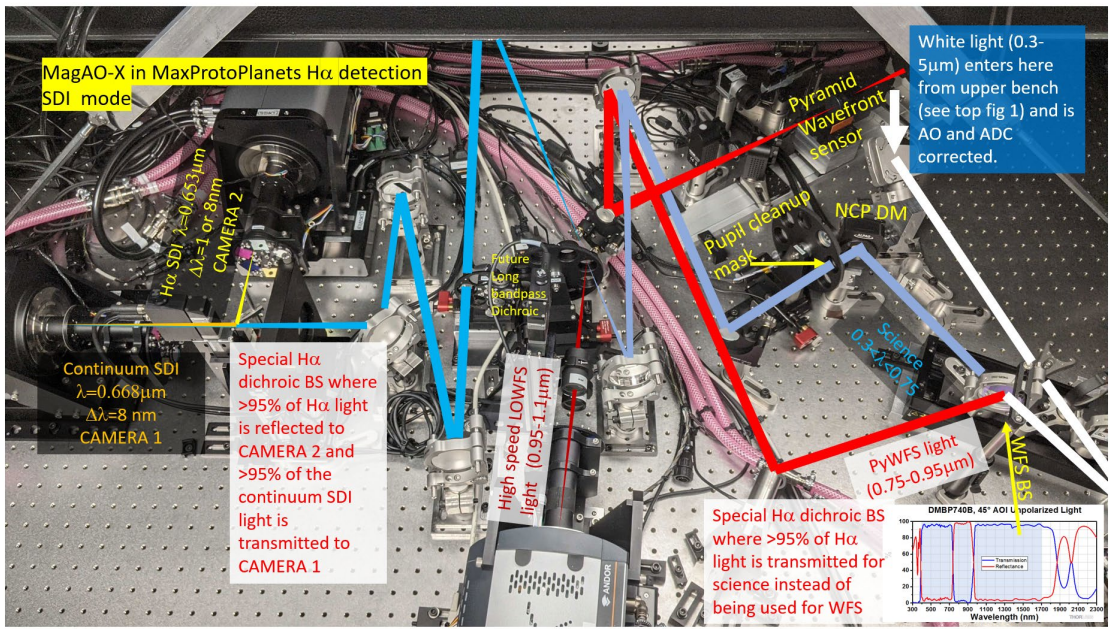

**Fig. 1:** Here we show the Hα SDI mode of the MagAO-X instrument used in this work (see Close et al. 2018 for a full optical description). This photo is of the lower optical table (circa 2022-2023). The upper bench (not shown) contains the 97 element woofer DM followed by the ADCs and the 2040 actuator Tweeter DM. Then a f/57 periscopic relay passes the AO and atmospheric corrected beam down onto the lower bench **(this is represented by the white arrow to the upper right of this figure).** Then the white light is collimated and passed through a special custom beamsplitter that transmits all the continuum and Hα light to the dual science cameras at f/69 shown to the far left. The rest of the light is reflected to the pyramid wavefront sensor (PyWFS). The bad edges (~3% undersized) of the primary and the one bad Tweeter DM actuator are blocked with the clean-up pupil "bump-mask" mask. In the Hα SDI mode shown above the beamsplitter optics are designed to optimize the total instrument atmosphere-to-CCD Hα QE to 16.6% at Hα.

Two different approaches could lead to substantial increases in the number of Hα planets detected. For bright (I <12 mag) targets, planets could be detected with MagAO-X and for those fainter (I >12 mag) with the laser guide star fed VLT/MUSE IFU. Hence, it is very important for the future of this field to know if the current lack of Hα detections is fundamental to the Hα line luminosity production (and extinction) mechanisms and/or variability –or simply a result of selection effects in the current generation of AO surveys (or a combination of both selection effects). We will test this with a survey of the best transitional disk targets for Hα protoplanets, we call this survey MaxProtoPlanetS. The obvious start to MaxProtoPlanetS is to use MagAO-X on PDS 70 and look for variability and ease of detection of these known protoplanets. These PDS 70 observations are a key "test-piece" observation for our new MaxProtoPlanetS survey and is the subject of this manuscript.

## 2.3. Introduction to PDS 70 b and c Past Detections

PDS 70 A is a 0.8 $M_{sun}$ TTauri star of age 5 Myr accreting at ~$6 \times 10^{-11} M_{sun}/yr$ (Thanathibodee et al. 2020), which has a spectacularly large 76 au wide disk gap (Keppler et al. 2019). Imaging with SPHERE was able to discover thermal emission from the circumplanetary disk and atmosphere from the gap planet PDS 70b (Keppler et al. 2018). We were able to use MagAO to discover Hα from PDS 70b (Wagner et al. 2018). The VLT's MUSE IFU was used to confirm the Hα emission from PDS 70b and discovered PDS 70c as another Hα protoplanet inside the gap (Haffert et al. 2019). Since, the separations of PDS 70b and c are rather large (~0.19" and ~0.23" respectively; circa 2018), telescopes like Keck at L' (3.8 μm) are able to follow-up these planets to detect their emission (Eisner 2015) where the masses are measured to be roughly ~2-4$M_{jup}$ for b and ~1-2$M_{jup}$ for c (Wang et al. 2020b).

We caution that while L' is superior to Hα to piercing any dust extinction, it can be hard to achieve the required spatial resolutions and inner working angle (IWA) at longer wavelengths. For example, PDS 70 b at 0.185" translates to just 2.5 λ/D at L' with the large D=10m Keck telescope, this is very close to the IWA limit for high-contrast (~$10^{-4}$) direct detection. Closer-in planets at, say, ~0.1" (1.4λ/D at Keck) would really require an ELT sized aperture for high-contrast direct direction at L'. In contrast, Hα is a 5.8x shorter wavelength, so even a smaller D=6.5m telescope finds a 0.1" planet at 5λ/D at Hα, and so can be detected, quite easily, (especially if there is use of a coronagraph) even if it is at $10^{-4}$ contrasts. Hence, there is a need for Hα surveys for close-in (≤0.1") protoplanet detection to compliment infrared observations that are more sensitive at wider separations.

## 3.0 MagAO-X OBSERVATIONS OF PDS 70 b AND c

### 3.1. The April 2022 Observations of PDS 70

The MagAO-X instrument was developed at the University of Arizona with an NSF MRI grant (Males et al. 2018). We partially commissioned MagAO-X in November 2019 and then shipped the instrument back to Tucson, Arizona and were planning to ship it back for April 2020 run, however, due to COVID the 6.5m Magellan Clay telescope was closed to visitors until our 2nd commissioning run in April 2022. Despite the long break from the telescope we were able to make major upgrades to the instrument in the lab. These upgrades included a high throughput Hα mode with all custom λ/10 beamsplitters (with ~95% transmission of Hα) where the Hα photons are transmitted to the science cameras and only ~5% are lost to the wavefront sensor optical path (see Figure 1 for details). Moreover, this mode also allows a very efficient SDI camera set up where another custom λ/10 beamsplitter cube transmits ~95% of the Hα continuum to a continuum filter ($\lambda_{CONT}$=668.0 nm; $\Delta\lambda_{CONT}$=8.0 nm) in science camera 1. This cube simultaneously reflects ~95% of the Hα light to an "wide" Hα filter ($\lambda_{H\alpha}$=656.3 nm; $\Delta\lambda_{H\alpha}$ =7.9 nm) to science camera 2. For the optical design of these two SDI EMCCD science cameras see the left hand side of Fig. 1.

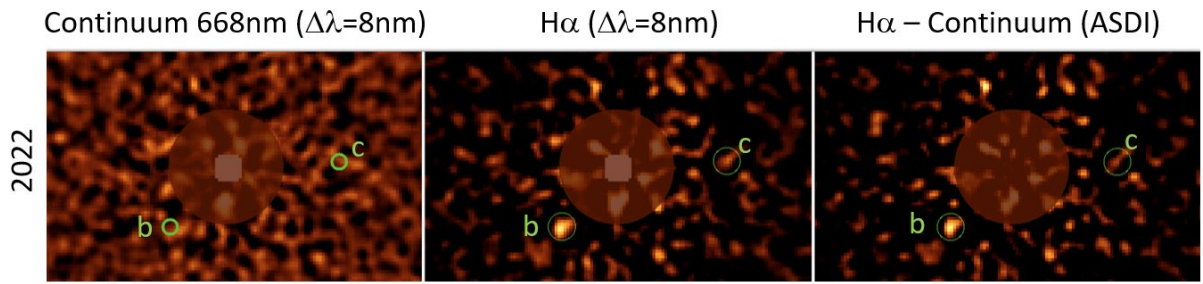

**Fig 2**: The pyKLIP reduced 2393 2s images of the April 2022 dataset. The pyKLIP parameters are 10 KLIP modes removed, movement=0, high pass filtered at 5.3 pix, and 1.3 hours of total integration (frames stacked into 39x 120s images that are fed into pyKLIP with 96° of rotation). The green circles (r=13mas continuum; r=20mas Hα and ASDI) have identical centers in all images and are centered on the predicted planet orbital location from the orbit of Wang et al. (2021). Following Mawet et al. (2014) the SNR of the ASDI b detection is 5.3 while c is weakly detected at just SNR~2.2. This image is 454.3x762.75 mas, and North is up, and East is left in these, and all following, images.

For clarity and completeness we list all the environmental, instrumental, and reduction settings in Table A1 in Appendix A for each night PDS 70 was observed. Please see Table A1 for a summary of all the settings and conditions of all our PDS 70 observations.

In our 2nd commissioning run in April 2022 we observed PDS 70 (with the Hα SDI mode shown in Fig. 1) in excellent seeing (0.4"-0.5") for 2.25 hours centered on the transit of PDS 70. We were able to lock the AO loop on the I=10.5 mag PDS 70 A with 460 modes at 666 Hz (we locked the PyWFS loop with very high EMgain of 600x "photon counting" mode of the OCAM2 PyWFS camera). The best 63.3% of those 2s integrations images at Hα had a FWHM of 29.5 mas (Strehl=8-12%), which is good –but still considerably worse than the diffraction limit of 21 mas at Hα (we note that the MagAO-X pupil clean up "bump-mask" hides the poor outer and inner edge of the M1 primary and makes the telescope effectively a D=6.31m scope; hence λ/D=21 mas at λ=656.3nm).

Since both the continuum (CONT) and Hα science cameras (cameras 1 and 2; respectively) are EMgain cameras (Princeton Pro EM CCDs) it is crucial that the EMgain is well calibrated from each camera. This was accomplished in the usual manner with dome flat field images (F1, F2) and zero (or bias) images (Z1, Z2) taken at the correct gain levels entered on the camera software ($EM_1$=$EM_2$=100 in 2022). Note that since these are EMCCDs there is an additional factor of √2 in the Poisson noise (so σ=$(2N)^{1/2}$ ) since the photon noise is also amplified. Hence, the gain equation becomes EMgain (ADU/e-) = "classical gain"/2; where "classical gain" = $(\sigma(F1 - F2))^2$-$(\sigma(Z1 - Z2))^2$/[<F1> + <F2> - (<Z1> + <Z2>)]. We find that in April 2022 that the EMgain_CONT = 24.22±0.14 ADU/e- (readnoise=0.92e- rms) in the individual 2s continuum images and EMgain_Hα = 35.46±0.03 ADU/e- (readnoise=0.48e- rms) in the Hα images. Later we will see that our measured Hα line flux depends on the ratio of EMgain_CONT / EMgain_Hα which was 0.683.

This ratio was confirmed on-sky by observing the well extended (hence a convenient "on-sky dome flat") cool hypergiant star VY Canis Majoris. VY CMa is a very flat Hα spectra source (see Fig. 4 in Humphreys et al. (2005)) again making it like a dome flat field (but it can be observed through the true f/69 instrument optical path at night). The ratio of the VY CMa fluxes = (Hα/CONT)=$\Delta\lambda_2/\Delta\lambda_1$*$QE_2$/$QE_1$*EMgain_Hα/EMgain_CONT
=(7.9/8.0)*(15.3/15.4)*EMgain_CONT/EMgain_Hα =0.98*(EMgain_Hα/EMgain_CONT)

From the observed ratio of the VY CMa ADU counts (Hα/CONT= 41235/28774) we can measure that EMgain_CONT / EMgain_Hα = 0.684; just 0.001 off the dome flat value of 0.683 which is excellent agreement given the ~1% uncertainty of the "flatness" of VY CMa spectrum around Hα. Hence, our dome flat gain measurements are verified on-sky, and we adopt EMgain_CONT = 24.22±0.14 ADU/e- and EMgain_Hα = 35.46±0.03 ADU/e- for the April 2022 observations.

The individual SNR of the planet detections were not high in this 2022 dataset compared to our 2023 and 2024 datasets. Nevertheless, we followed the procedure of Mawet et al. (2014) to calculate the SNR for each planet. We used an aperture of r=FWHM (30mas) and calculated average flux of planet b in that aperture (this is the signal; S). We then calculated the standard deviation (noise; N) of the average fluxes in each of r=FWHM aperture at the planet b distance from A. We had 16 completely independent apertures at r=158 mas (making a ring of independent noise apertures around PDS 70 A at the radius of planets b). We then applied the correction of Mawet et al. (2014) and found a SNR of 5.3 for planet b in Fig. 2. For planet c the same procedure was followed and a SNR of ~2.2.

We put very little statistical weight on our "detection" of c in 2022. In our 2023 and 2024 datasets the detections of c are much more significant, not because c was brighter, but because we optimized the MagAO-X instrumental set-up (we all this an instrumental "contrast boost"; see Table A1) for detection of faint Hα planets after our 2022 commissioning run.

The MagAO-X coronagraphs were not utilized for any of our PDS 70 observations as the target star brightness (r'=11.65 mag) was low enough that little speckle noise actually contaminated the Hα PSF at the location of the planets (see Appendix A; Fig. A1); hence it was overall more important to maximize Hα throughput (since we were photon-starved of planet Hα photons) and so avoid any Lyot stop throughput losses. For brighter (r'~6 mag) MaxProtoPlanetS targets; like HD100456, we have successfully used the MagAO-X Lyot coronagraph to increase contrasts.

### 3.2. The March 2023 Observations of PDS 70

In our 2nd science run in March 2023 we observed PDS 70 in good seeing (0.45"-0.55") for 5 hours starting 3 hours before the transit of PDS 70. Even before we slewed to the target we switched into the Hα SDI mode (as shown in Fig. 1) and locked the PyWFS on a bright I=6 mag guide star (close to PDS 70's position in the sky) and engaged the low-order WFS mode (LOWFS) of the 2 science cameras. The LOWFS uses phase diversity to measure any non-common path (NCP) aberrations (McLeod 2023). Once these LOWFS NCP errors were measured we used our 97 actuator Alpao DM-97 NCP DM to remove NCP from both camera's f/69 focal planes. These NCP errors evolve very slowly (MagAO-X is floating and gravity invariant; Close et al. 2018) and so we can use a bright star near PDS 70's coordinates to optimize the NCP DM into the right shape to minimize the NCP errors. The resulting Hα images were excellent and high Strehl on the bright calibration star were obtained.

We then froze this NCP DM shape and made a small telescope offset to the location of PDS 70 where we locked the PyWFS loop with very high EMgain of 600x "photon counting" mode of the OCAM2 PyWFS camera (<0.6e- rms noise). We were able to control 536 modes at 1 kHz for 4 hours (within ±2 hours of transit). For over 4 hours we had >95% of the 2s Hα and continuum images acquired had FWHM<28 mas and a very stable PSF. The final combined 2023 image was roughly double the Strehl (20% vs. 9%) of the previous April 2022 commissioning data (despite the slightly worse seeing). This improved performance is due to the better PyWFS AO response matrix calibration and the use of the LOWFS to remove the static NCP errors from the science cameras.

Dome flats taken in March 2023 when the cameras were set to $EM_1=100$ and $EM_2=300$ were reduced to find $EMgain\_{CONT} = 24.21\pm0.12$ ADU/e- (readnoise=0.92e- rms) in the individual 2s continuum images and $EMgain\_{H\alpha} = 102.13\pm0.09$ ADU/e- (readnoise=0.16e- rms) in the Hα images. So in 2023 $EMgain\_{CONT}$ / $EMgain\_{H\alpha} = 0.237$ . This ratio was confirmed on-sky by observing the well extended cool star VY CMa in the 2 cameras. The VY CMa observations confirmed the dome flat ratio $EMgain\_{CONT}$ / $EMgain\_{H\alpha} = 0.237$ within <2% error.

Another significant improvement in the March 2023 data was the use of a special custom Alluxa 656.3 nm Hα filter that had 95% throughput but only a Δλ=1.045 nm. This ultra-narrow Hα filter allows the same amount of planet Hα photons to be detected while minimizing the amount of continuum starlight that otherwise "leaks into"/contaminates wider Hα filters. The exact "contrast boost" from the commissioning run set-up in 2022 to our optimized SDI Hα mode in 2023 can be estimated from the β parameter (explained in section 6). From Table A1 we see $β_{2022}/β_{2023}$=7.27/0.87= 8.4x (where β is defined in table A1). This implies that an Hα planet that decreases in flux by 8.4x would be detected at the *same* Hα_contrast in 2023 (with our high EMgain and 1nm filter) as a planet with *no* decrease in flux but observed in the sub-optimal 2022 set-up.

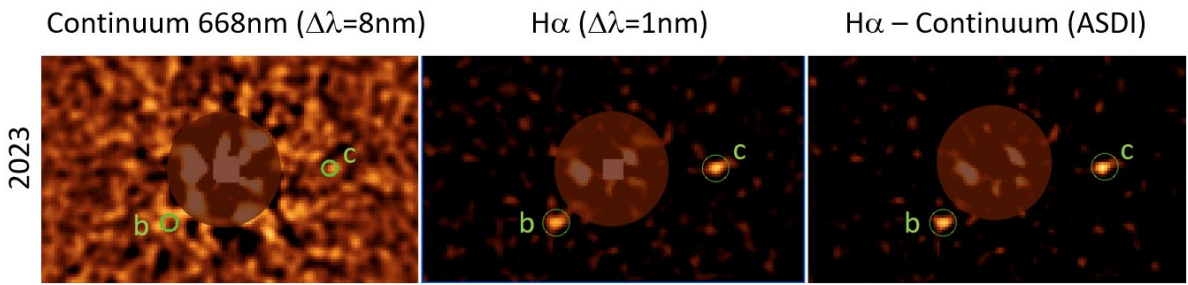

**Fig 3:** The pyKLIP reduced 6573 2s image March 2023 dataset. Data and pyKLIP parameters are 10 KLIP modes removed, movement=0, high pass filtered at 5.3 pix, 3.6 hours of integration (219x 60s images fed into pyKLIP with 137° of rotation). The green circles are identical centers in all images. Following Mawet et al. (2014) the SNR of the ASDI b detection is 10.4 while c is very well detected at SNR=13.1. This image is 454.3x762.75 mas in the x and y directions.

We selected a symmetric 3.6 hour period centered on the transit of PDS 70. This gave a continuous sampling from -68 deg of parallactic angle to +69 degrees (so 137 degrees of total rotation). We selected the 96.7% of the data that had FWHM<28 mas and so we had 3.6 hours of total integration (6373 2s frames) simultaneously for both the Hα 1nm filter and the 8nm 668nm continuum filter. The detailed reduction of these data is outlined in section 4.

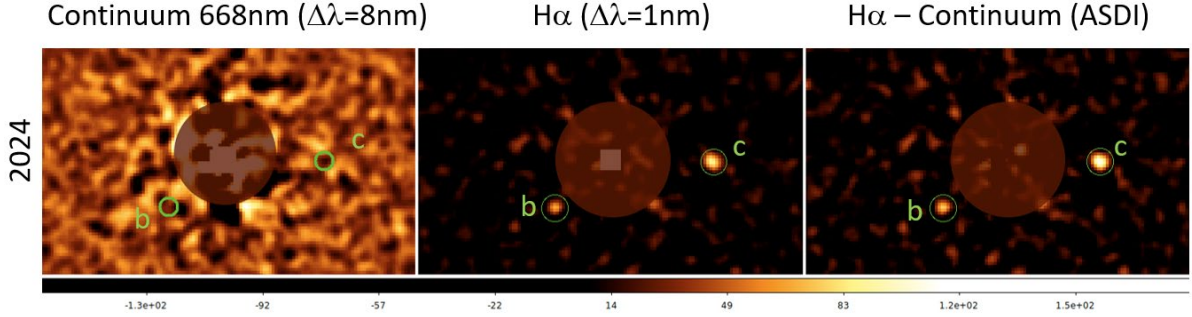

**Fig 4:** The pyKLIP reduced 7124 1s image March 2024 dataset. Data and pyKLIP parameters are 10 KLIP modes removed, movement=0, high pass filtered at 5.3pix, 2 hours of integration (118x 60s images fed into pyKLIP; 89° of rotation). The green circles have identical centers in all images and are centered on the predicted planet orbital location from the orbit of Wang et al. (2021) at this epoch. Following Mawet et al. (2014) the SNR of the ASDI b detection is 4.3 while c is well detected at SNR=12.3. Each of the images are 454.3x762.75 mas in size.

### 3.3. The March 2024 Observations of PDS 70

There was even more improvement in our 25 March 2024 dataset shown in Fig. 4. The March 2024 dataset was similar to the March 2023 dataset (see Table A1). However, a key difference was that the NCP DM had been upgraded from a Alpao DM97 to a 1024 actuator BMC 1K DM, this allowed for much better removal of non-common path aberrations. These NCP errors were measured and removed using a bright I=6 mag star with FDPR phase diversity just before the PDS 70 dataset was taken (Van Gorkom et al. 2021). Also the optical throughput was slightly increased by 1.14x (compared to the past observations) to a total QE of 16.6% in H$\alpha$ (QE=16.8% in continuum) by removing the "bump mask" from the pupil. But the ratio $QE_{CONT}/QE_{H\alpha}$ remained constant at 1.01 for all years (see Table A1). This ratio is actually the only QE term that the H$\alpha$ line flux depends on –as we will see later when the $\beta$ parameter is introduced in section 6. These throughput measurements were based on photometric standards in photometric conditions at airmass ~1.

Dome flats were taken in March 2024 when the cameras were set to $EM_1$=200 and $EM_2$=600. We measured the $EMgain_{CONT}$ = 45.84±0.47 ADU/e- (readnoise=0.48e- rms) in the individual 1s continuum images and $EMgain_{H\alpha}$ = 196.09±0.17 ADU/e- (readnoise=0.16e- rms) in the H$\alpha$ images. So in 2024 $EMgain_{CONT} / EMgain_{H\alpha}$ = 0.234 . This ratio was confirmed on-sky by observing the well extended cool star VY CMa in the 2 cameras. The VY CMa observations confirmed the dome flat ratio EMgain_CONT / EMgain_Hα = 0.234 within <1% error on sky.

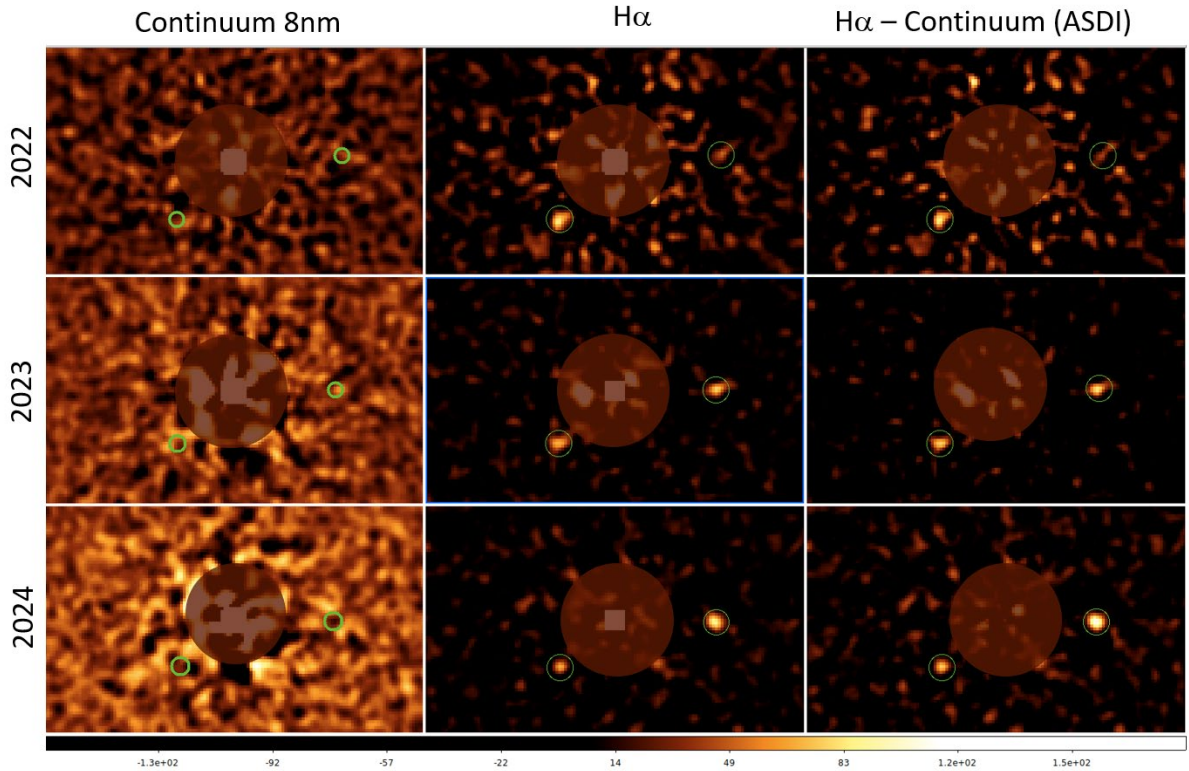

**Fig 5:** Summary of all the datasets with the parameters from figures 2, 3 and 4. Note how much better c has been detected in 2023 and 2024 vs. 2022. We get this "contrast boost" in 2023-2024 by: use of the narrow Δλ=1nm Hα filter (which passes all the Hα planet light but minimizes starlight); better AO correction; better readnoise; and better removal of the non-common path errors. This all results in the Hα Strehl improving from 9% in 2022 to 20% in 2023 and still further to 26% in 2024.

## 4.0 REDUCTIONS

Data reduction was designed around the fact that the flux from these planets at Hα is very low, indeed we only expect approximately ~3 Hα planet photon to be detected by a given pixel every minute (the pixels are very small at just 5.9 mas/pix). This implies that one needs to average 30x 2s exposures together before there is a good chance of >1 detected planet photon per pixel within one FWHM of the planet core. Therefore a custom python/pyIRAF pipeline was developed that optimized the preservation of individual photon events while also maximizing the contrast with ADI and SDI (which we call ASDI).

*4.1. The New MaxProtoPlanetS Low Hα Flux Pipeline*

The first step in the pipeline was the selection of the highest Strehl data. This has already been mentioned in the above section, but to optimize the Strehl and the total integration time 63% of the 2022 raw frames were kept (for a total of 1.3 hours of integration; 8-12% Strehl) and in 2023 the better AO correction allowed 96.7% of the data to be kept for 3.6 total hours of integration (6573x 2s selected images in a narrow range of 15-25% Strehl) and 86.6% of the frames were kept in 2024 for 2 hours of total integration (7124x 1s frames in a range of 22-30% Strehl).

The pipeline then takes the selected images and removes any >7σ cosmic rays or >7σ EM CIC noise and replaces those pixels with bias values. Then the peak of the PDS 70 A star is located and a 256x256 subsection (1.46x1.46") is removed (centered on PDS 70 A). Then we use the pyIRAF *xregister* cross-correlation function with the spline3 interpolator to allow all the frames to be shifted to with ±0.01pix of a reference PSF image (of PDS 70 A from this dataset) that has been shifted to be centered exactly at 128.00, 128.00.

The IRAF spline3 interpolator was found to best preserve individual "photon" events which are significant for these observations without leading to overshoot (like we, unfortunately, found with the SINC31 interpolator once EMgains>150 ADU/e-). As is shown in Fig 6 (middle and right of top row) and also in Fig. A1 (Appendix A), at the locations of planets b and c we are effectively "photon starved" from the stellar PSF (with <0.3 star ph/pix in a 2s exposure >150 mas from the star) hence we are operating in a "Geiger" mode where we are detecting individual photon events at the r>150mas locations of the planets. The flux rate from the planets themselves is ~3-4x lower than the stellar PSF at ~0.1 planet Hα ph/pix in a 2s exposure. Hence each photon event detected must be preserved by the pipeline. We found that the spline3 interpolator worked very well to preserve these individual planet photon events but eliminated ringing or overshoot when interpolating them.

At this point the pipeline has produced 6573 aligned frames in the, for example, 2023 dataset. We then "average by time" and reduce these 6573 frames into 219 groups of 30 frames and summing 30x 2s frames together to make a 219x 60s exposures. We find that each of these 219 frames have, on average, ~3 Hα planet photons detected on each pixel within the FWHM of the planet (r=2.5pix=FWHM/2) so integrating over a r=2.5pix aperture yields ~60 Hα planet photons/min in our r=FWHM/2 detection aperture. This is enough planet signal per 60s image for pyKLIP to fit the speckles ~10x fainter than this (~6 photons/speckle) given the 219 individual 60s ADI input frames to KLIP. Hence, it is possible to reveal the planets with SNR>10.

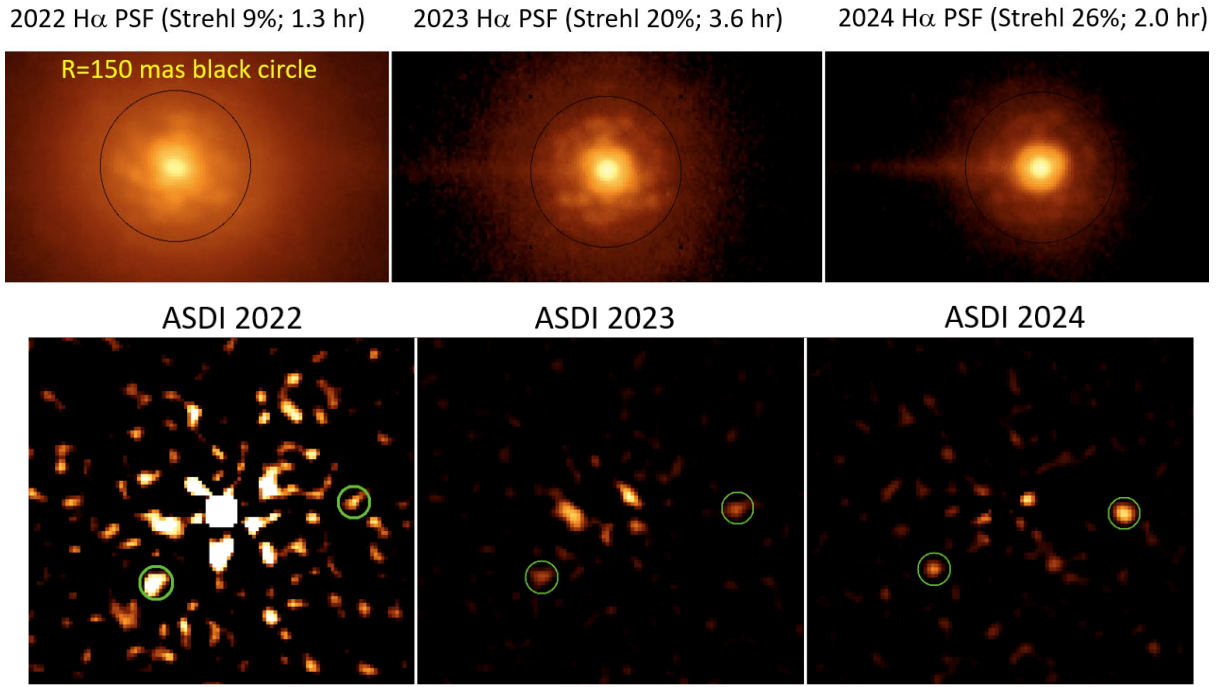

**Fig 6:** The top row shows the Strehl ratio of the long term Hα PSF for each dataset (see Appendix A for our methodology to measure Strehl). Note how the amount of PSF light dramatically fades quickly at r=150mas (where b is located) as Strehl increases. The bottom shows the final ASDI images (from Figs. 2-4) but with a stretch to represent the observed Hα line flux. Hence, in 2022 planet b has a much higher intensity appearance then it does in 2023 where its line flux falls by 4.6x. Planet c is clearly 2.3x brighter in 2024 compared to 2023. Variability of the line flux of both planets is obvious and significant.

We have found experimentally that this ~3 planet photon/pix is the minimum amount of signal at ~155 mas (location of b) to have the planet signal optimally survive the pyKLIP process with good SNR. Our pyKLIP tests with 438x 15s images with just ~1.5 ph/pix had lower SNR. Whereas, pyKLIP tests with 109 images of 120s each, also had lower SNR --possibly due to have too few KLIP images and too much rotation (~1.5 deg between 120s images). In any case, the best SNR was obtained by the pipeline summing 30x 2s images to produce a final set of 219x 60s images.

Next the pipeline high-pass filters these 219 images. This removes the otherwise "hard-to-scale" stellar halo from each 60s image. We found the SNR reasonable with a 5.3 pix Gaussian filter for the 2022 data. Hence, we standardized on 5.3 pix for all the years, so the datasets could be compared to each other. This process of high-pass filtering the data also effectively removes any low spatial frequencies that remain from the bias or flat field. As a result we do not need to carry out classical flat field or bias subtraction, and so we minimize excess random noise that would otherwise be introduced by those steps. We do take flat fields for gain calibration and we find that the EMCCDs are flat (in the 256x256 area used) with <2% of flat pixel-to-pixel QE variation, hence this flat noise cannot add significant noise as we are continuously rotating the planets over different CCD pixels (as the sky rotates w.r.t. our detector). Therefore any flat error will average to zero, so our choice of not flat fielding is justified.

The last step in this stage of the pipeline is to accurately fit a radial profile to each of the 60s high-pass filtered PSFs. This PSF profile is stored and used later by pyKLIP for the planet PSF in the forward modeling "planet injection" part of the pipeline. To maximize the contrast of the planets w.r.t. the background (and the speckles that we want to fit and remove) we need to first remove any remaining low spatial frequencies. The biggest remaining source of low frequency power is the radial profile of the PSF itself. Hence, we simply subtract the fit radial profile from each 60s image, removing most of the azimuthal power in the PSF, but leaving the speckles and planets largely untouched.

### 4.2. pyKLIP

We then used standard PCA starlight ADI removal by the popular pyKLIP (Wang et al. 2016) package to reduce the 219x 60s frames in the usual fashion. We found there were many combinations of pyKLIP parameters that gave similar SNR detections of the planets. We did not try to search all pyKLIP parameters to aggressively maximize SNR of each planet as that can add positive bias to the planet fluxes. In fact, the b and c planets had similar SNR with 1, 5, 10, 20 and 40 KLIP modes removed giving us strong confidence in these detections. Also the number of pyKLIP sectors and annuli did not greatly matter either (we adopted 4 sectors and 10 annuli for all reductions in this work). The pyKLIP movement parameter did have an noticeable effect in that smaller movements (0-2) lead to some self-subtraction of the planets, but removed most of the extended PSF and disk contamination (so a good choice for point-sources like b and c); whereas

movement=5 was good for disk detection. Very large movements (10-15) left too much PSF/disk residuals. In the end, we chose a movement of 0 for the pyKLIP reduction of the 2022 dataset (see Fig. 3) because this maximized both removal of the strong stellar halo which was a real issue as Strehl was only 9% in 2022 (see upper left figure in fig 6). We then decided, for consistency, to also use movement 0 for the 2022 and 2023 reductions as well. So, to be clear, there was no attempt to run through a large grid of pyKLIP parameters to optimize the very best SNR at the planet locations --as this can artificially increase the flux of the planet detected by conflating speckle noise and planet flux (see Follette et al. (2023) and Adams Redai et al. (2023) for a review on pyKLIP parameter selection and exploration).

## 5.0 PHOTOMETRY AND ASTROMETRY

As PSF subtraction algorithms (like pyKLIP) can distort planet signal, **so** we obtained companion astrometry and photometry through the Bayesian KLIP Astrometry (BKA) technique with the forward modeling feature in PyKLIP (Wang et al. 2015) for accurate measurements and uncertainties on the planet contrasts.

With BKA our approach to photometry and astrometry of planets b and c was straightforward. We used the fully forward modeled planet insertion option of pyKLIP to inject fake negative planets at the separations of b and c into each of the 219 input images. We fit a Gaussian to the final multi-hour "deep" PDS 70 A PSF to find an accurate stellar peak counts and planet FWHM to model the fake planets as accurately as possible. This, in turn, leads to the most accurate contrasts astrometry from BKA.

But we first needed to remove the stellar speckles and disk residuals from the H$\alpha$ KLIP reduced image as well as possible. To do this we scale the continuum image (which has no H$\alpha$ emission in it) to the ratio of the $\text{StarFlux}_{\text{H}\alpha}$ / $\text{StarFlux}_{\text{CONT}}$. Subtracting this scaled continuum image from the H$\alpha$ image removes most (but not all) of the stellar and disk residual speckles and creates the ASDI image (see right hand side of Figs. 2-5).

We used standard Gaussian PSF fitting photometry tools to estimate the rough locations of the planet's center of light in the ASDI image (see far right Fig 3). A grid of fake forwarded modeled negative planets (starting with a slightly too negative fake planet) were then injected with $\Delta X = \pm 2.5$pix, $\Delta Y = \pm 2.5$pix around the planet's center of light in 0.5 pixel steps (100 steps total). Once the position of the planet is well established (by a symmetric dark hole at what was once the previously positive planet's position) a grid of fake negative planet contrasts from $1\times10^{-3}$ to $1\times10^{-4}$ were injected in steps of $1\times10^{-5}$. If the planet subtraction looked asymmetric (due to a small centering error) we saved the current value of the planet flux and redid the astrometric grid (but just $\Delta X=\pm1.0$ pix, $\Delta Y=\pm1.0$ pix), and then finished the contrast search. We continued to make planet subtracted ASDI images until the flux inside a r=FWHM aperture (centered on the planet) fell to zero in the ASDI image after fake negative planet injection.

However, there is still residual photon noise and speckle noise in the ASDI images. We need to measure the standard deviation of the speckle noise at the radii of the planets. We accomplished this by inserting fully forward BKA modeled planets with the exact same brightness of the planet, but at various different position angles. See Appendix B Fig. B1 for an image of our "ring of fake planets" from which the error in our contrasts from the residual noise is measured. For example, we found that for the 2024 dataset the Hα contrasts (ASDI image upper left Fig. B1) at which the b planet was subtracted to ~zero residual planet flux (upper right Fig B1) integrated over a r=FWHM aperture was $ASDIcontrast_{H\alpha}$ = $(4.7\pm1.1)\times10^{-4}$. The contrast error bars were determined from the forward modeled "ring of N fake planets" which had a mean flux of: mean$\pm[\text{sum(flux-mean)}^2/(N-1)]^{0.5}$ = $(5.3\pm1.1)\times10^{-4}$ for b and $(6.1\pm0.50)\times10^{-4}$ for c. So we adopt an relative error on the $ASDIcontrast_{H\alpha}$ of b of 23% and 8.1% for c for the 2024 epoch (see appendix B for full details on the error analysis). So in 2024, c has actually surpassed b in brightness ($6.1\times10^{-4}$ vs $5.3\times10^{-4}$), as can be clearly seen by inspection of Fig. 5.

| **EPOCH: 24 April 2022** | **PDS 70 b** | **PDS 70 c** |
|---|---|---|
| **Observed Separation** (mas) | **158.1±3.0** | **218.3±5.9** |
| **Observed PA** (deg) | **135.5±0.5** | **272.0±0.5** |
| Predicted orbital Separation (mas) Wang et al. (2021) | 159.787±1.157 | 224.24±2.257 |
| **Error** (Obs. Sep. – Predicted orbital Sep) (mas) | -1.7 ± 3.2 | -5.9± 6.3 |
| Predicted orbital PA (deg) Wang et al. (2021) | 135.452±0.325 | 272.064±0.240 |
| **Error** (Obs. PA – Predicted orbital PA) (deg) | -0.05 ± 0.60 | +0.06 ± 0.56 |
| Forward modeled contrast results from photometry | | |
| **ASDIcontrast$_{H\alpha}$** : (flux of planet in ASDI image) / (Hα flux of star) | **(2.8±0.4) x 10$^{-4}$** | **(0.9±0.4) x 10$^{-4}$** |
| **Hα line flux of planet** $f_{H\alpha}$ (erg/s/cm$^2$) | (10.4±1.6) x 10$^{-16}$ | (3.3±1.5) x **10$^{-16}$** |

**Table 1:** The Photometry and Astrometry for April 2022 for PDS 70 planets b and c. Values in **bold** text are directly measured, otherwise they are calculated values.

| EPOCH: 8 March 2023 | PDS 70 b | PDS 70 c |
|---|---|---|
| **Observed Separation** (mas) | **157.5±3.0** | **206.5±1.0** |
| **Observed PA** (deg) | **132.18±0.50** | **270.00±0.25** |
| Predicted orbital Separation (mas) Wang et al. (2021) | 155.106±1.021 | 209.821±1.10 |
| Error (Obs Sep – Predicted orbital Sep) (mas) | +2.4±1.0 | -3.3± 1.5 |
| Predicted orbital PA (deg) Wang et al. (2021) | 132.581 ± 0.324 | 270.020 ± 0.240 |
| Error (Obs PA – Predicted orbital PA) (deg) | -0.40±0.59 | -0.02 ± 0.35 |
| Forward modeled contrast results from photometry | | |
| **ASDIcontrast$_{H\alpha}$:** (flux of planet in ASDI image) / (H$\alpha$ flux of star) | **(4.75±0.45) x 10$^{-4}$** | **(4.25±0.32) x 10$^{-4}$** |
| **H$\alpha$ line flux of planet** $f_{H\alpha}$ (erg/s/cm$^2$) | (2.28±0.26) x 10$^{-16}$ | (2.04±0.21)x 10$^{-16}$ |

**Table 2:** The Photometry and Astrometry for March 2023 for PDS 70 planets b and c. Values in **bold** text are directly measured, otherwise they are calculated values.

| EPOCH: 25 March 2024 | PDS 70 b | PDS 70 c |
|---|---|---|
| **Observed Separation** (mas) | **150.5±3.0** | **206.55±3.0** |
| **Observed PA** (deg) | **130.18±0.50** | **268.0±0.5** |
| Predicted orbital Separation (mas) Wang et al. (2021) | 149.770±1.707 | 208.000±1.060 |
| Error (Obs Sep – Predicted orbital Sep) (mas) | +0.79±3.45 | -1.45±3.16 |
| Predicted orbital PA (deg) Wang et al. (2021) | 128.909 ± 0.526 | 267.474 ± 0.356 |
| Error (Obs PA – Predicted orbital PA) (deg) | 1.27±0.73 | 0.52 ± 0.61 |
| Forward modeled contrast results from photometry | | |
| **ASDIcontrast$_{H\alpha}$:** (flux of planet in ASDI image) / (H$\alpha$ flux of star) | **(4.7±1.1) x 10$^{-4}$** | **(6.2±0.5) x 10$^{-4}$** |
| **H$\alpha$ line flux of planet** $f_{H\alpha}$ (erg/s/cm$^2$) | (3.64±0.87) x 10$^{-16}$ | (4.78±0.46)x 10$^{-16}$ |

**Table 3:** The Photometry and Astrometry for March 2024 for PDS 70 planets b and c. Values in **bold** text are directly measured, otherwise they are calculated values.

An identical procedure was followed for the 2022 and 2023 datasets. In 2022 we find for b ASDIcontrast$_{H\alpha}$ = (2.8±0.4)x10$^{-4}$ and for c (which was only marginally detected at the SNR~2.2 level) ASDIcontrast$_{H\alpha}$ = (0.9±0.4)x10$^{-4}$. In 2023 we find for b ASDIcontrast$_{H\alpha}$ = (4.75±0.45)x10$^{-4}$ and for c ASDIcontrast$_{H\alpha}$ = (4.25±0.32)x10$^{-4}$.

For the astrometry we observed an astrometric calibration field in Baade's Window which has been extensively used by MagAO and GPI. This field gave an astrometric solution for Camera$_2$ of 0.00589±0.00004"/pix, PA$_{offset}$=+2.1±0.2 deg platescale in 2022 and 0.00590±0.00004"/pix, PA$_{offset}$=+2.0±0.2 deg in 2023 (and assumed for 2024) values adopted from the averages given in (Long et al. 2024).

In addition to the contrasts listed above we report the astrometry in Tables 1-3. A key result from the astrometry of Tables 1-3 is that the 2022, 2023 and 2024 positions of b and c Model-Obs. errors (lines 4 and 6 in each of Tables 1-3) are consistent (≤2σ) with zero, hence the model orbit of Wang et al. (2021) is well followed the Hα planets. This is the most definitive proof, to date, that the Hα emission region is coincident with the thermal photosphere tracked interferometrically by VLTI/GRAVITY in the orbit of Wang et al. (2021). We can now state, with some confidence, that the Hα emission has an origin within r<1.7mas for b and r<2.5mas for c and PA<0.7º (1.9mas) and PA<0.4º (1.4 mas) for c from the thermal center --since these are standard deviation of the mean of the Obs-model values. This is an error ellipse of 0.2x0.2 au around b and 0.28x0.16 au for c (semi-major axis in radial direction to star). This supports the magnetospherical accretion theory for the generation of the Hα (see the accretion model of Thanathibodee et al. (2019)) but it still allows for compact accretion shocks onto the circumplanetry disk above the planet as well (see, for example, Szulágyi et al. 2022).

## 6.0 ANALYSIS

### 6.1. Example Hα Line Luminosity Calculation for PDS 70 b in 2024

The $L_{H\alpha}$ luminosity can be calculated for a gap planet of an extinction corrected effective "r' mag" at Hα (which we call r'mag_p_Hα) by comparing the its flux with Vega:

$$L_{H\alpha} = 4\pi D^2 f_{H\alpha} = 4\pi D^2 vega_{zero_{point}} \Delta\lambda \{10^{\frac{r'mag\_p\_H\alpha}{-2.5}}\} \quad (1)$$

Where $f_{H\alpha}$ is the Hα line flux and r'mag_p_Hα is just the effective de-extincted "r' magnitude" w.r.t. Vega for planet "b" at Hα. This is good approximation since the effective center of the r' filter is close to that of Hα. Moreover, the center of the R=658 nm filter is exactly at Hα; and we find PDS 70 A's flux is very similar at R and r' (R-r'=0.04 mag from the UCAC4 2012 catalog). It is clear that r'mag_p_Hα is related to the observed r' mag of the star (r'$_A$ = 11.65±0.06 mag; UCAC4) minus the common extinction to both the star and the planet ($A_R$). There is also the possibility that there is extra extinction towards the planet ($A_p$) in addition to $A_R$. We cannot easily determine $A_p$, yet Zhou et al. (2021) estimate that since PDS 70 b is well detected in the UV the extinction $A_R+A_p$< 3 mag. However, for the sake of comparison to other values in the literature we will assume no extinction to the planet ($A_R = A_p = 0$) even though we suspect some extinction ($A_p$) is likely, but we lack a convincing way to measure extinction from just a single emission line (measuring the ratio Hβ/Hα would help solve this, but Hβ has proved elusive to measure; Haffert et al. 2022). There have also been efforts (unsuccessful to date) with infrared AO to detect Paβ emission from b or c (see Uyama et al. (2021) for example). However, Uyama et al. (2021) do estimate $A_V$~0.9 and ~2.0 mag for b and c; respectively. Still literature flux values typically assume no extinction, so we do as well for sake of comparison, but clearly these line fluxes are actually lower limits as some extinction ($A_R + A_p$ is somewhere in range 0-3 mag; Zhou et al. 2021) is quite possible around these dusty young planets.

There is also a slight correction for the leakage of the primary's continuum into A's Hα measurement, which causes ΔmagHα (the contrast in just the Hα image, no SDI) to be slightly larger than it should be. This can be completely removed by using the ASDI contrast (ASDIcontrast$_{Hα}$ = flux of the emitted Hα photons from the planet line (with no continuum contamination; ASDI image) divided by the flux of the star at Hα.

It is then necessary to tie the photometric system from the Hα flux of PDS 70 A (which is too variable) to the continuum flux of PDS 70 A; so we to need calculate: ASDIcontrast$_{continuum}$ = Flux_Hα/StarFlux_Cont . We need to compare to the 668nm continuum since it is steady with time (indeed we found only a 10% increase/yr in the absolute continuum flux of PDS 70 A from 2022 to 2024; see Table A1). To solve for ASDIcontrast$_{continuum}$ from our observables takes a few steps:

Since the # of planet Hα photons:

Flux_Hα = ASDIcontrast$_{Hα}$ * StarFlux_Hα , therefore:

ASDIcontrast$_{continuum}$=Flux_Hα/StarFlux_Cont=ASDIcontrast$_{Hα}$*StarFlux_Hα / StarFlux_Cont

But we need to convert the StarFlux from ADU to photo electrons (e-) so:

ASDIcontrast$_{continuum}$ = ASDIcontrast$_{Hα}$*StarFlux_Hα / StarFlux_Cont * EMgain_CONT/EMgain_Hα * QE$_{CONT}$/QE$_{Hα}$

Which is awkward, so we can introduce a parameter β so that:

$$\text{ASDIcontrast}_{continuum} = \text{ASDIcontrast}_{Hα} * β$$

Where, β = StarFlux_Hα / StarFlux_Cont * EMgain_CONT/EMgain_Hα * QE$_{CONT}$/QE$_{Hα}$

Where all of the parameters of β are easily measured ratios (all are listed in Table A1). The fact that β is completely dependent on ratios minimizes systematic errors which simply divide out in each ratio. We estimate that the error is less than 2% in β comparing dome flat gains to on-sky measured gains with VY CMa observations. In 2024 StarFlux_Hα / StarFlux_Cont =0.816 and EMgain_CONT/EMgain_Hα = 0.234 and QE_CONT/QE_Hα = 1.01; therefore, β= 0.193 in 2024.

Therefore, we can use the above relation to write equation 2:

$$\Delta magASDI_{continuum} = 2.5*\log_{10}(ASDIcontrast_{continuum}) = 2.5*\log_{10}(ASDIcontrast_{H\alpha} * \beta) \quad (2)$$

Since ASDIcontrast_Hα is (4.7±1.1) x $10^{-4}$ (Table 3), therefore from equation 2 we know ΔmagASDI_continuum is 10.10±0.25 mag in 2024. There is also a very slight correction since there is extra ~0.05 mag added due to Hα light in r' filter mag. So, the "r' mag" of the planet is:

$$r'mag\_p\_H\alpha = (r'_A - A_R) + (\Delta magASDI_{continuum} + 0.05) - A_p = (11.65 \pm 0.06 - A_R) + (10.10 \pm 0.25 + 0.05) - A_p \quad (3)$$
$$= 21.8 \pm 0.26 \text{ mag}$$

In the case of PDS 70b we have very little extinction to the star and so we will assume $A_R=A_p=0$ (Wagner et al. 2018; Thanathibodee et al. 2019; Zhou et al. 2021) so equation 3 suggests an effective r'mag_p_Hα flux of b at Hα is similar to a continuum 668nm (Δλ=8nm) source with an r'~21.8 mag flux. Therefore, the line luminosity $L_{H\alpha}$ can be written:

$$L_{H\alpha} = 4\pi D^2 Vega_{zero_{point}} \Delta\lambda \{10^{\frac{r'mag\_p\_H\alpha}{-2.5}}\} \quad (4)$$

Which we can directly solve for in the case of PDS 70 b as:

$\log(L_{H\alpha}/L_{sun}) = \log(4\pi(113*3.1\times10^{18})^2 * 2.4\times10^{-5} * 0.008)/[3.9\times10^{33} * 10^{((21.8\pm0.26)/2.5)}] = -6.84$

where the Vega zero point magnitude of the r' filter (Vega_zeropoint; Males 2013) is 2.4x$10^{-5}$ ergs/(s cm² μm). Since we are comparing the Hα flux to that of PDS 70 A in the continuum filter we use Δλ=0.008 μm for our continuum filter. This $\log(L_{H\alpha}/L_{sun})$ = -6.84 is a significant amount of emission, but significantly less than before 2023 as we will see later.

To calculate the Hα line flux ($f_{H\alpha}$) of b is simple, just divide the line luminosity $L_{H\alpha}$ by $4\pi D^2$ as is clear from equation 1. Therefore, the Hα line flux can be shown to be (3.64±0.87)x$10^{-16}$ erg/s/cm² with a full and correct Gaussian error analysis of equation 4. See Appendix B and Fig. B2 for our Gaussian error propagation to produce the distribution of line flux's for b in 2024. In this manner all the Hα line fluxes and errors in the last row of Tables 1-3 were calculated.

## 6.2. Mass Accretion Calculation for PDS 70 b and c

Since low mass, young, objects have excellent Xshooter calibrated accretions rates (Rigliaco et al. 2012) we can use:

$$L_{acc} = 10^{[2.99\pm0.23 + (1.49\pm0.07)*(\log(L_{H\alpha}/L_{sun}))]} \qquad (5)$$

from the empirical total accretion luminosity $L_{acc}$ to $L_{H\alpha}$ relations of Rigliaco et al. (2012) for very low mass accretors. This formula may not apply at these low masses and accretion rates. Indeed, Thanathibodee et al. (2019) find, by applying the first full treatment of H$\alpha$ line radiative transfer in a magnetospheric geometry for planetary-mass objects, that weakly accreting planets accreting below the cut off of $\dot{M}_p<10^{-12} M_{sun}/yr$ are better fit with a $\log(\dot{M}_p)$ varies as $0.353\log(L_{H\alpha}/L_{sun})$ power-law, so the production of H$\alpha$ is less efficient. Hence, we find that in this case this a power-law:

$$L_{acc} = 10^{[-3.62+0.353\log(L_{H\alpha}/L_{sun})]} \qquad (6)$$

better describes the relationship between H$\alpha$ luminosity and total luminosity for low planetary accretion rates. So equation 6 clearly yields lower estimate accretion luminosity ($L_{acc}$) for the weakest accretors. Then using the standard relation (equation 7) relating the released total accretion luminosity $L_{acc}$ from accretion onto the planet surface:

$$\dot{M}_p = 1.25 L_{acc} R_p/(GM_p) \qquad (7)$$

of Gullbring et al. (1998), yields a planetary accretion rate estimate given by equation 7. By using $M_p$ mass estimate of ~4 $M_{jup}$ for PDS 70b (Wang et al. 2020b) and a planet radii ($R_p$) from the 5 Myr COND evolutionary model (estimate of $R_p=1.3R_{jup}$; Baraffe et al. 2003) then in the case of b using (5) and (7) yields $\dot{M}_p=1\times10^{-13}$ $M_{sun}/yr$. Using equation (5) and (7) for c we find $\dot{M}_p=1.2\times10^{-13}$ $M_{sun}/yr$ assuming a mass of ~2 $M_{jup}$ for PDS 70c (Wang et al. 2020b) and a radius of $R_p=1.3R_{jup}$ (Baraffe et al. 2003). So in 2024 it is clear that the mass accretion is similar for b and c at around $\dot{M}_p\sim1\times10^{-13}$ $M_{sun}/yr$ for both planets, which is much less than the cutoff of $\dot{M}_p<10^{-12}M_{sun}/yr$ found by Thanathibodee et al. (2019) with equations (5) and (7) so we instead should apply the low accretion rate formula of equations (6) and (7) for both planets. That yields $\dot{M}_p=1.2\times10^{-12}$ $M_{sun}/yr$ for b and $\dot{M}_p=1.4\times10^{-12}$ $M_{sun}/yr$ for c. We can also write this as $\dot{M}_p\sim6\times10^{-5}$ $M_{jup}/Myr$ which is a very low rate and might represent the end of the planetary growth period. This is logical given the rather old ~5 Myr age of the system at which point gas is disappearing from the disk and the gas giant growth period is ending.

All our calculations are approximate due to uncertainty in the correct forms of equations 5 and 6. Moreover, this work assumes no dust extinction toward the planets (to be consistent with other works in the literature). It is likely that there is some dust extinction to the source of the H$\alpha$ emission. While the *HST* UV excess images b by Zhou et al. (2021) do suggest that $(A_R+A_p)<3$ mag for b --we need still to be cognitive that, if the extinction is ~3mag, then the "true" de-extincted line fluxes would be ~16x greater. We fully explore this impact on the range of possible $\dot{M}_p$ values in Appendix D and figure D1. Fig D1 shows that the order of magnitude of the accretion is $\dot{M}_p \sim 1\times10^{-12}$ $M_{sun}$/yr which is still quite weak accretion compared to the mass of the planets. It appears that the PDS 70 planets are currently accreting gas at a low rate compared to their masses.

NOTE: All of these mass accretion values are correct in this preprint, but in the published version of this paper there was a 24x scaling error (mass accretion rates were too low by 24x) for some of these $\dot{M}_p$ values in section 6.2. The $\dot{M}_p$ values in this preprint should be used instead.

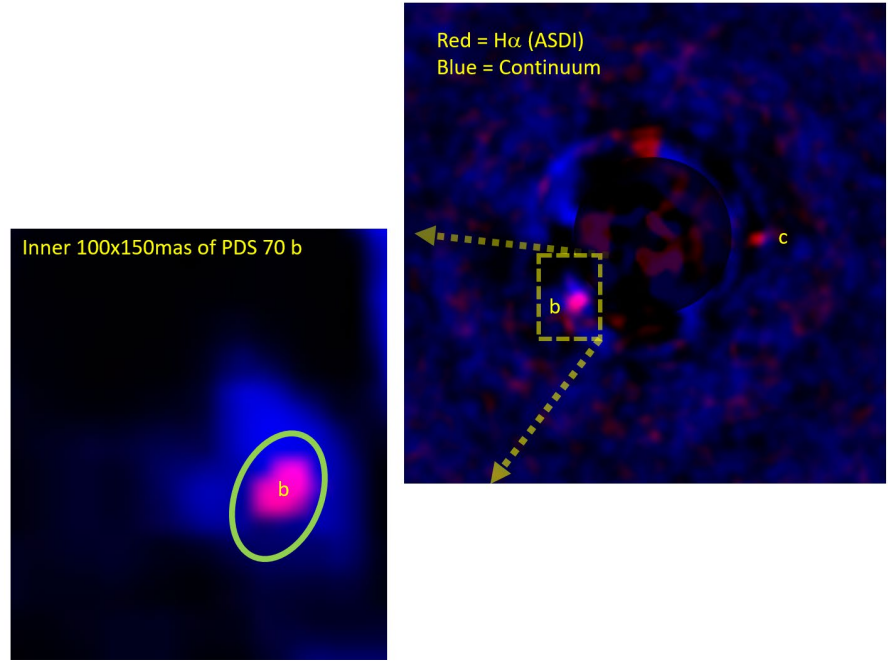

**Fig 7:** Possible scattered light off the PDS 70 b CPD. Here we see in an 0.94x0.94" 2-color image that the continuum image from 2023 (shown in blue) shows a strong **scattered light** feature centered on the Hα emission from PDS 70 b (shown in pink). We see that the Hα emission of b is that of a point source. We note that the slight elongation in the radial direction of the Hα emission is typical of a point source with some self-subtraction with the pyKLIP reduction. This pink point source subtracts very well with a forward modeled planet as noted in the photometry section.

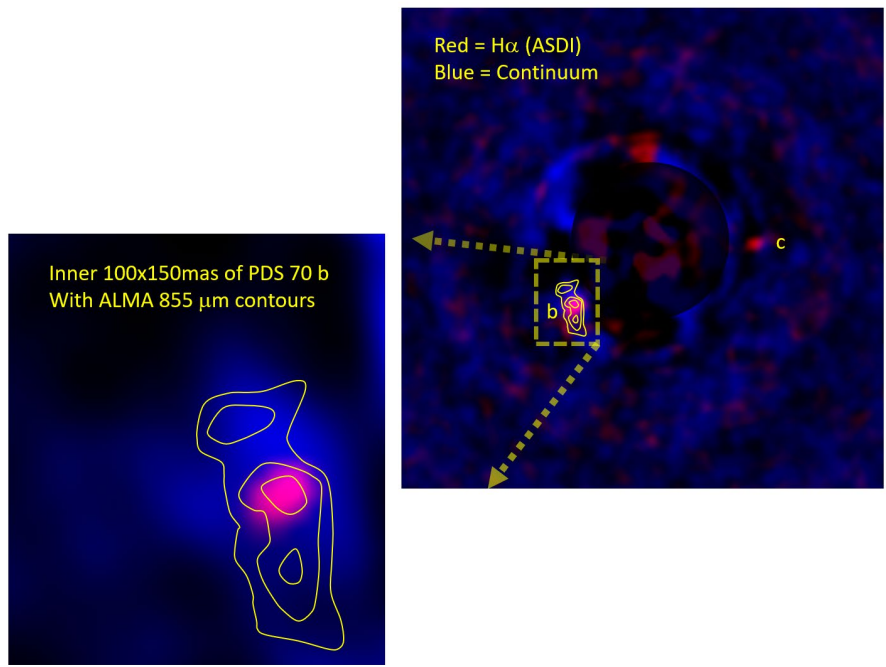

**Fig 8:** Here we see that the continuum image from Fig 5. (shown in blue) shows a strong elongation that is roughly consistent with the circumplanetry dust detected by O. Balsalobre-Ruza et al. (2023) with ALMA at 855 μm (yellow contours). Note we have superimposed these contours at the peak of b's emission assuming the 855μm dust is orbiting around PDS 70 A with b.

# 7.0 DISCUSSION

## 7.1. Reflected Light from Circumplanetary Dust around PDS 70 b and c

An important, and novel, result from this work is the clear detection of reflected light from the immediate environment surrounding PDS 70 b. As is clearly seen from Figs. 7-9 there is an extended (but very compact) continuum "ring-like" dust continuum structure with FWHM~60 mas (~40 mas in 2023 and ~70 mas in 2024) with the Hα emission point sources located in (or near) the center (the planets would be at the center of these "dust ellipses" if looking at a deprojected view of a flared disk). This continuum emission is the brightest flux detected at r≥0.16" separations from the star in 2023 and 2024 and so is very unlikely to be a nearly identical PSF noise artefacts for 2 years in a row, it is more likely that we are witnessing (for the first time) scattered light off a real dust ring around planet b (2023 & 2024) and planet c in 2024. This could be a circumplanetary dust disk (CPD) centered on the planets with signs of light scattering off dust in the disk "front" side closest to the star.

We believe this resolved scattered light circumplanetary feature has not been noted before in the optical, which is not surprising since no other observations to date obtained 2-3.5 hours of ~25-27 mas images in a dedicated continuum filter at such blue wavelengths. However, there is some evidence that ALMA has detected low levels of dust emission with a similar size and azimuthal angle shown in figure 8. Indeed the combined 855μm ALMA dataset of O. Balsalobre-Ruza et al. (2023) clearly shows a dust structure around b (what they label as $b_{ext}$) which is both similar in orientation and size as the dust structure in Fig. 7. We super impose the ALMA contours reported by O. Balsalobre-Ruza et al. (2023) onto our continuum image in Fig 8. *Figs. 7-9 illustrate that there is good evidence of compact circumplanetary dust around the PDS 70 protoplanets.*

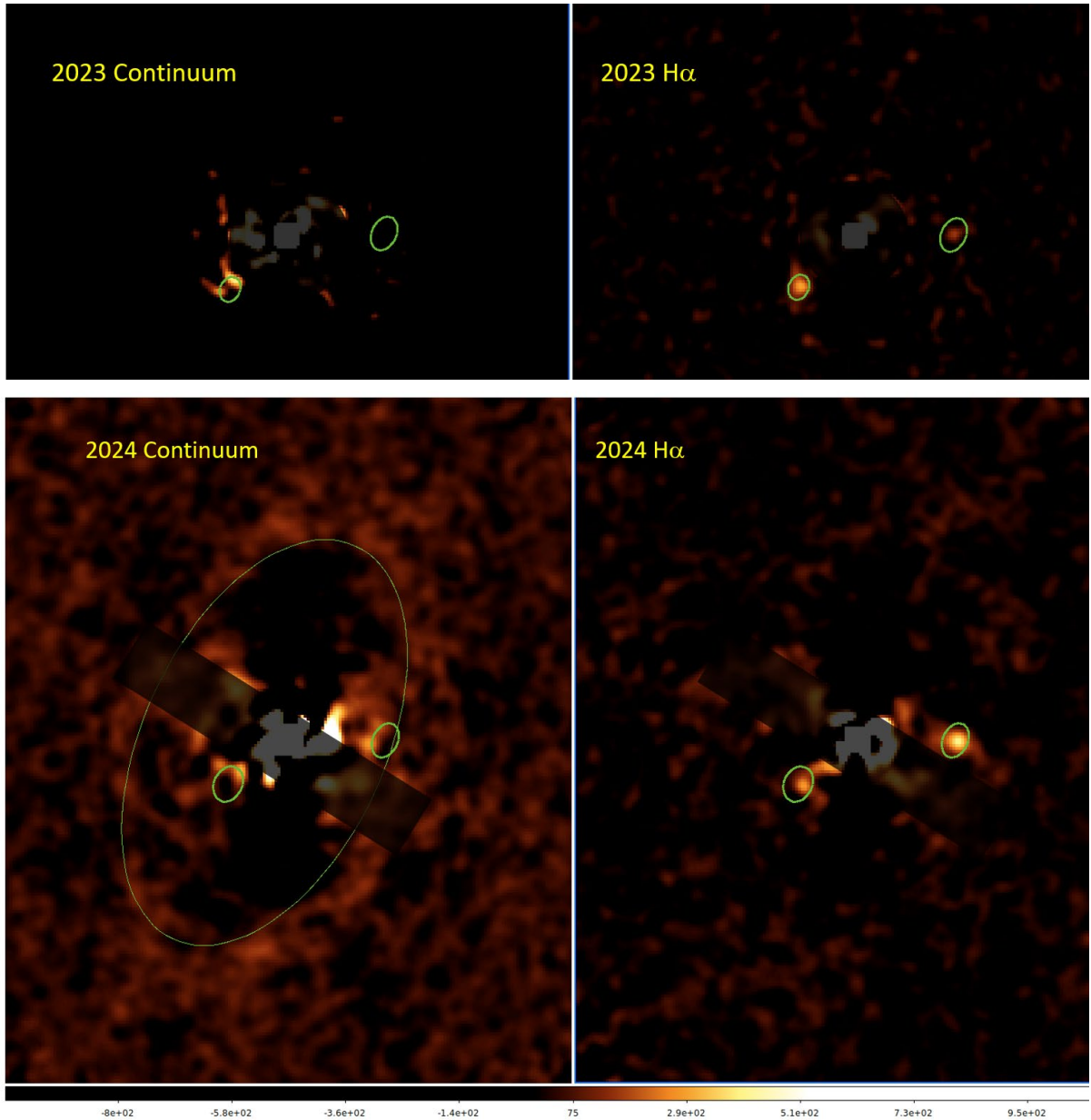

**Fig 9:** Here we compare the continuum and Hα. Reduced the same as in figs 3-4, except pyKLIP movement=5 (instead of 0), and a high pass filter at 19.5pix (instead 5.3pix), with these settings the fainter extended disk features are preserved. In the 2024 continuum image (lower left) we see light scattered of the PDS70 transition disk, with the dark cavity fit (r=49 au; 0.437 mas) with the thin green ellipse. We scale down this ellipse and center them on b and c in the Hα images. We then copy identical ellipses to the continuum images. Size of these ellipses are set by the peak illumination at the front of the CPD. We see this "bright spot" in the continuum at ~20mas from b in 2023 and ~35mas in 2024. This suggests an average disk of radius of ~30 mas around the b planet (or $r_{disk}$~3 au when deprojected). We do not detect a disk around c in 2023 (nor in the 2022 data which too low quality), but in 2024 there is a "bright arc" feature that is well fit by the long edge of disk (see the "c" green ellipse in lower left image). Hence, c might also have ~3.5 au disk (deprojected). Lower left FOV is 1.14x1.38".

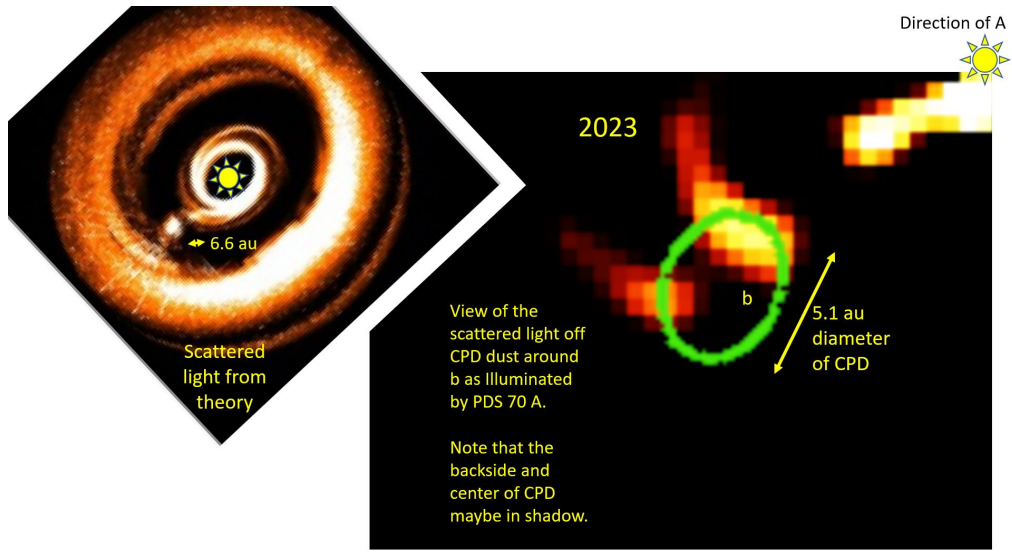

Fig 10: **Left:** The radiative transfer scattered light model of Szulágyi & Garufi (2021) show that for a 10 $M_{jup}$ protoplanet at 20 au one would expect a "CPD" or "dust torus" diameter of ~6.6 au (2 Hill Sphere radii) reproduced from fig A1 of Szulágyi & Garufi (2021) with infalling dust. **Right:** The observed dust distribution (continuum filter) in 2023 around b is in reasonably good agreement with the predicted scattered light model of a CPD. The CPD has a diameter of ~5 au and is brightest on the edge facing into the starlight with the back and center of the flared CPD possibly in shadow.

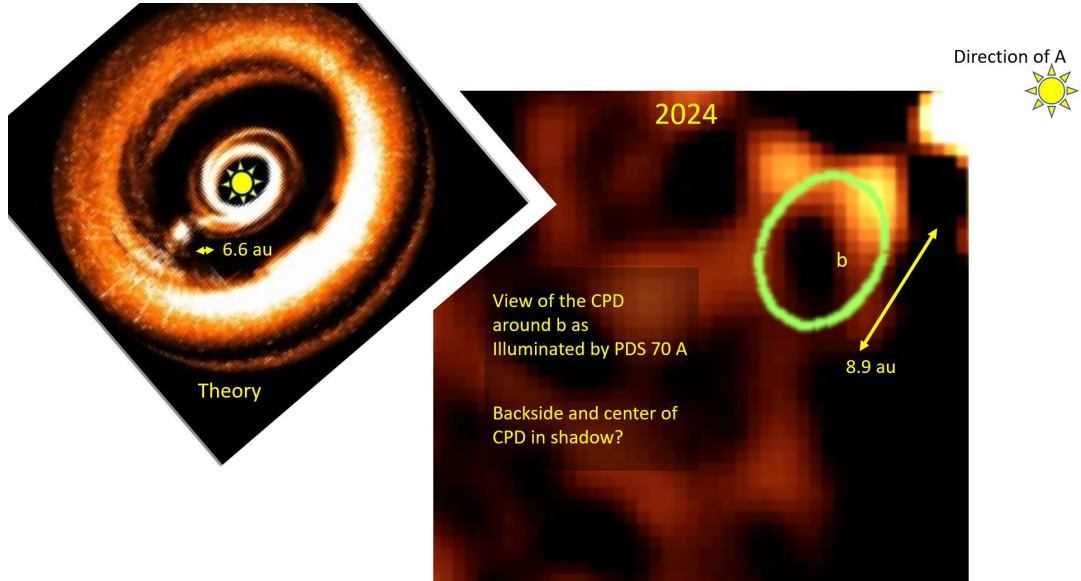

**Fig 11**: Similar to fig 10, but for the 2024 dataset. The 2024 observed dust distribution around b is reasonably good agreement with the scattered light radiative models of Szulágyi & Garufi (2021). Where the size of the CPD diameter or "dust torus" is estimated at ~9 au. The back and center of the flared CPD may be in shadow from the brightly illuminated front side edge.

The size of this disk is $r_{disk}$ ~30 mas (~3 au deprojected) this is much too large to be the "classical thin, high density" inner classical "hot" circumplanetary disk which is expected to be 1/3-1/2 that of the Hill sphere (see, for example, Szulágyi et al. (2022) and references within). However, the very detailed 3D radiative hydrodynamic models of both an accreting protoplanet and circumstellar disk of Szulágyi et al. (2022) show that one expects there to be a highly flared circumplanetary disk (CPD) with an accretion shock surface on the top from the incoming mass flux from the circumstellar disk. If one was to then illuminate this CPD surface from PDS 70 A one would expect the edge of the disk facing the star to be brightest from Mie scattering (to appreciate this scattering geometry see our cartoon in Fig. E1). Indeed, as Figs. 9 and E1 make clear, the CPD feature between b and the star is the brightest feature that we find at r≥150 mas from the star in 2023 and 2024. Hence, we have good reasons to believe that these bright disk features are real, and to some degree, expected from models. For example, the models of Szulágyi & Garufi (2021), and the equation for the Hill-sphere, predict that for a protoplanet at 20 au (like PDS 70 b) one would expect the radius of this disk to be r~2.5 au (assuming 4 $M_{Jup}$ planetary mass; see Figs. 10-11). Indeed we observe a slightly larger $r_{disk}$~2.5-3.5 au (deprojected) which is a consistent with a higher planetary mass in the range of 4-12 $M_{Jup}$; however, the higher end of these mass regime can be ruled out from the non-detection of the planet itself in continuum. In any case, all this infalling small grain size dust once within the Hill-sphere is orbiting around the planet. For PDS 70 b (and c) we also expect the Hill sphere ($R_H$) to be $R_H$~3 au (from equation 1 of Close 2020), so the observed r~3 au dust is likely bound (at least temporally) to the planets and is effectively the outer part of the planet's CPD. *These might be the first images of CPDs in scattered light imbedded inside a larger circumstellar disk (see Fig E1),* and should help inform future planet formation models. Future observations will be helpful to increase our confidence in the size of these CPDs, which could provide good constraints on the planetary mass.

However, it is possible, but unlikely, that this very bright "dust disk" morphology is due to an "unlucky" repeating spike of speckle noise reproduced in 2023 for b and in 2024 for b and c. At this point it is more likely that these are real scattered light features from CPD dust disks around the protoplanets; nevertheless, further observations would be helpful to confirm their nature. In particular, a future study will properly forward model theoretical disks to see if these pyKLIP images match theory (with accurate self-subtraction), but this is beyond the scope of this manuscript

## 7.2. The Variability of the PDS Protoplanets Over 7 Years

We can look at the Hα line flux of the protoplanets as a function of time. There have been several successful attempts to image the planets b and c over the last 7 years. In figure 12 we show how the Hα line flux from planets b and c have varied with time. While there are likely some unknown systematics between flux values in the literature and MagAO-X, we know the MagAO-X 2022-2024 data all used the same instrument and mode, and so errors estimates are reliable and can be intercompared between those datasets.

| Reference | Telescope/ Instrument | Obs. Date | Separation of b (mas) | PA of b (deg) | Hα filter Flux of b $\times 10^{-16}$ erg/s/cm$^2$ | ASDI Hα line flux b* $\times 10^{-16}$ erg/s/cm$^2$ | ASDI Hα line flux c* $\times 10^{-16}$ erg/s/cm$^2$ |
|---|---|---|---|---|---|---|---|
| Follette et al. 2023 | 6.5m Magellan/ MagAO | 8 Feb 2017 | -- | b not detected | -- | -- | 152.5±76 |
| Wagner et al. 2018 | 6.5m Magellan/ MagAO | 3,4 May 2018 | 183±18 193±12 | 148.8±1.7 143.4±4.2 | 33±18 | 34±14 | Not detected |
| Haffert et al. 2019/ Hashimoto et al. 2020 | 8.2m VLT/MUSE | 20 June 2018 | 176.8±25 | 146.8±8.5 | 8.1±0.3 | 8.1±0.3 | 4±2 |
| Zhou et al. 2021 | 2.4m HST/WFC3 | 2 Feb-3 July 2020 | 177.0±9.4 | 143.4±3.0 | 16.2±2.2 | 10.5±1.4 | Not detected |
| This work | 6.5m Magellan/ MagAO-X | 24 April 2022 | 158.1±3.0 | 135.5±0.5 | 16±2 | 10.4±1.6 | 3.3±1.5 |
| This work | 6.5m Magellan/ MagAO-X | 8 March 2023 | 157.5±3.0 | 132.18±0.50 | 3.0±0.2 | 2.28±0.26 | 2.04±0.21 |
| This work | 6.5m Magellan/ MagAO-X | 25 March 2024 | 150.5±3.0 | 130.18±0.50 | 4.4±0.2 | 3.64±0.87 | 4.78±0.46 |

**Table 4:** The astrometry of PDS 70 b and the photometry of PDS 70 b and c since discovery.
*zero extinction ($A_R=A_P=0$ mag) is assumed in all line fluxes, table modified from Aniket et al. (2022).

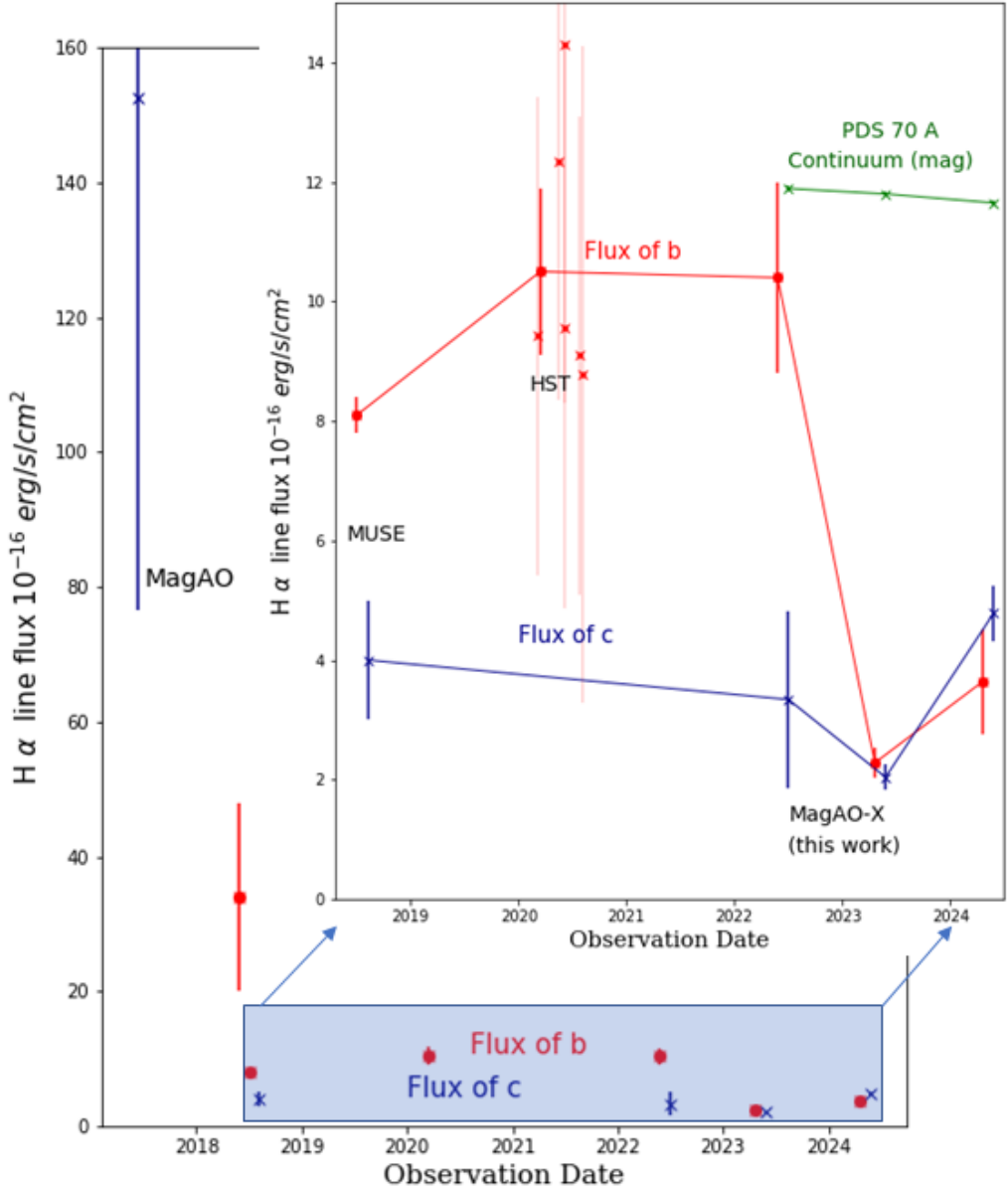

**Fig 12:** The calculated Hα line fluxes of planets b and c assuming no extinction ($A_R=A_P=0$ mag) from last 2 columns of Table 4. The main plot covers 7 years of observations and has a large Y axis upper limit to accommodate the January 2017 archival MagAO image recovery of c (Follette et al. 2023). Even though the errors are large, it appears that both c and b may have been much brighter at Hα around 6-7 years ago. Then from April 2018 to April 2022 planet b was roughly constant at $\sim 11\times10^{-16}$ ergs/s/cm$^2$ (see inset for zoom in) but then from March 2023 onwards b dramatically faded to just $\sim 2$-$3\times 10^{-16}$ ergs/s/cm$^2$. The planet c's line flux has been roughly constant since April 2018 at $\sim 5\times 10^{-16}$ ergs/s/cm$^2$. Yet, we see evidence that c's flux is changing with a slight (but significant) 2.3x increase from 2023 to 2024. The green points are the magnitudes of PDS 70 A that we observed in the continuum filter and converted to r'. This variability of PDS 70 A in the continuum was accounted for and calibrated out of our MagAO-X line fluxes in this work. These data suggest b is generally fading at Hα but c has recently brightened and now surpasses b.

It was important that any contamination from the continuum was removed from the literature datapoints in Fig. 12. This only affects the *HST* dataset for b as they just used the 2nm Hα filter which would have let in some continuum, this is particularly true in that the size of the continuum dust structures (see Fig. 9) are quite compact (<80mas) and since the *HST* Hα PSF is ~53mas (compared to the 27 mas in Figs 5-7) it would have been impossible to separate the Hα photons from the contamination of some extra continuum light. We have found that the continuum adds about 27% of the photons captured in a 54mas diameter photometric aperture in a 1nm Hα filter in our 2023 dataset. The *HST* observations of Zhou et al. (2021) were in a 2nm wide Hα filter and so our model suggests we should divide the flux by a factor of 1+0.27*(2nm/1nm) = 1.54x. That correction is applied to the *HST* photometry for planet b in the second to last column in Table 3 and is shown in Fig. 12. *HST* did not detect planet c so it is just the *HST* b fluxes in Fig 12 that needed this adjustment. All other datapoints in Fig. 12 are ASDI and so the continuum contamination is not a serious issue.

The main plot in Fig. 12 covers 7 years of observations and has a large range in the Y axis to accommodate the 8 February 2017 archival MagAO image recovery of c (Follette et al. 2023). Even though the errors are large for both the MagAO datapoints circa 2017-2018 those errors (which we computed with equation 4 and the error propagation outlined in Appendix B) are based on proper KLIP forwarded modeled ASDI contrast errors.

It appears that both c and b may have been much brighter at Hα around 6-7 years ago. It is also possible that the brightness of b and c in the 2017-2018 datapoints are contaminated by a bright random noise speckle that falls within the positions of b and c. While possible, this is unlikely because the c dataset had >90 degrees of rotation and 2 hours of integration (Follette et al. 2023) and the b data of Wagner et al. (2018) had 2 datasets with over 4 hours of integration -- and in all cases the seeing was 0.5" or better. However, with only ~120 modes well corrected by MagAO on a star as faint as PDS 70 A the Strehls were very low and the images had FWHM~50 mas, hence there was considerable PSF speckle residuals at the positions of b and c so KLIP had a significant task to dig out those MagAO detections, yet these are the oldest Hα detections of the planets and provide our only link to how bright the planets were ~6-7 years ago.

The planet b's flux appears roughly constant from April 2018 to April 2022 at ~11x10$^{-16}$ ergs/s/cm$^2$ (see Fig 12 inset) but then from March 2023 through March 2024 b dramatically faded to just ~2-3x10$^{-16}$ ergs/s/cm$^2$ . For example, in April 2022 b was (10.4±1.6) x10$^{-16}$ ergs/s/cm$^2$ and then fell to $(2.28\pm0.26) \times 10^{-16}$ ergs/s/cm$^2$, so a decrease in b's flux of $(8.1\pm1.6)\times10^{-16}$ ergs/s/cm$^2$ which is a very significant drop. Indeed, b fell by a factor 4.6x in flux from 2022 to 2023 and it continued to be faint in 2024 as well --slightly rising to $(3.64\pm0.87)\times10^{-16}$ ergs/s/cm$^2$ in March 2024. This drop in flux was recorded while our absolute photometry showed PDS 70A changing by less than ~0.1x year to year (r'=11.89, 11.80 and 11.65 mag, from 2022, 2023, and 2024) and while our photometry for c was also nearly constant. *Therefore, we conclude that this strong drop in b's Hα line flux is intrinsically from PDS 70 b itself, it cannot be due to some unknown systematic error that only affects b but not c or PDS 70 A. This is the first unambiguous detection of a change in Hα line flux from a protoplanet.* Characterizing such accretion variability can inform our models of how (and where) material is accreted onto protoplanets. Continued monitoring of the planets will be helpful in this effort.

The planet c's line flux has been roughly constant since April 2018 at $\sim 5\times10^{-16}$ ergs/s/cm$^2$. Yet we see evidence that c's flux is also changing, albeit on a smaller scale than for b. We measure a slight (but significant) increase of $(2.74\pm0.51)\times10^{-16}$ ergs/s/cm$^2$ from 2023 to 2024 (this is a 2.34x increase in flux). This is proof that c has increased significantly in flux from 2023 to 2024. It will be interesting to continue to observe c and see if it continues to increase in brightness.

It is also worth noting that the trend of b getting fainter and c getting brighter has led to c in 2024 being slightly $(1.14\pm0.97)\times10^{-16}$ ergs/s/cm$^2$ brighter than b (see Appendix B, Fig B2 to compare complete flux error distributions between b and c in 2024). These are the first observations to show c brighter than b. This is also further evidence that the fluxes of these protoplanets are significantly changing over the 2022-2024 period.

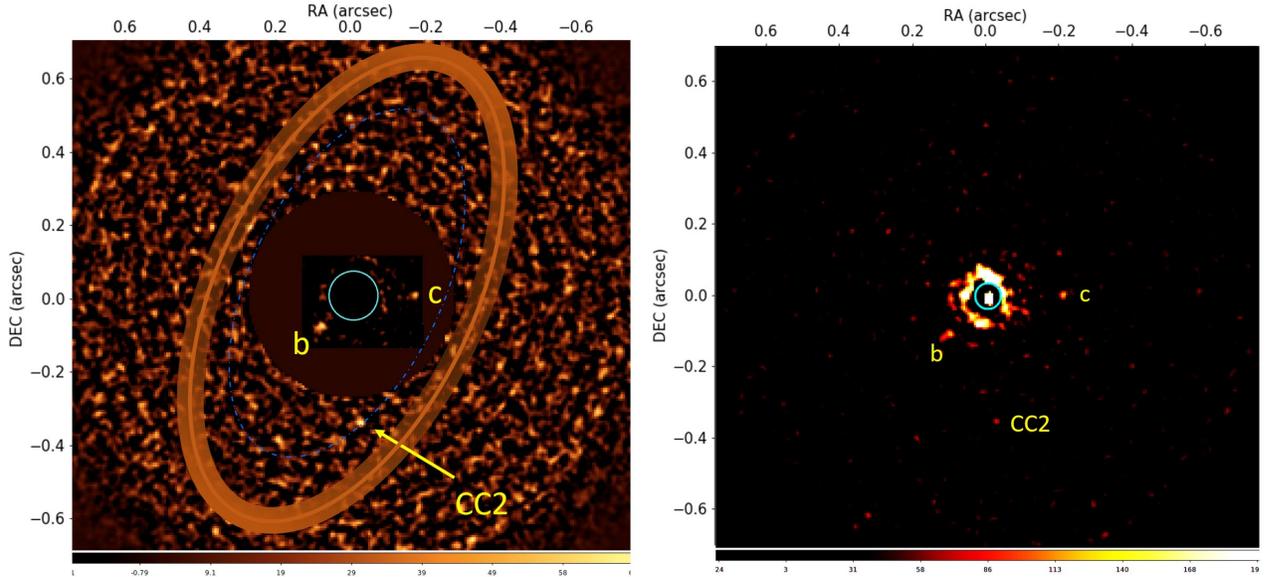

**Fig 13: Left:** A composite image of the outer 1.48x1.40" ASDI field with a classical ADI reduction around PDS 70 lacks any obvious Hα emission point sources besides c and b. The inner dark hole has a ~10x suppressed stretch and is 0.227" in radius. We did detect one marginally significant (~3σ) source (CC2 at sep=0.348" PA=184.0 deg) that seemed to be present in all the reductions in the 2022 and 2023 datasets. The outer ellipse is the rough position of the ALMA dust free cavity (r=76 au) and the faint dashed blue line is the possible co-planar orbit for CC2 (r~57 au)  **Right:** Same field with a single uniform stretch, CC2's ASDI contrast is ~7.5x10$^{-5}$ at ~3σ. Each of the sources as bright as CC2 were found to be strong continuum sources as well, only CC2 was the only outer source (other than b and c) significantly detected in Hα but not continuum. Unfortunately, CC2 was not detected in the 2024 dataset casting doubt on its existence.

*7.3. The Search for Other Outer Planets in the PDS 70 System: Candidate CC2*

Given the excellent datasets that we obtained it is logical to carry specialized data reductions aimed at the detection of faint outer planets. We carried out an extensive suite of different reductions with different pyKLIP parameters and CIC and cosmic ray rejection algorithms, and high pass filters. In the end there were no other highly significant Hα emission point sources found in any of our datasets. Due to the lack of stellar speckles past ~0.3" in the 1nm Hα filter we found the most sensitive data reduction approach was classical ADI (CADI) with an SDI subtraction of the continuum images from the Hα images, so same reduction as before but CADI instead of pyKLIP. The results of this reduction can be seen in figure 13.

In figure 13 there is no clear sign of any other planets in the PDS 70 system.  We did detect one source (which we call CC2; there was an unrelated "CC1" already found by *HST;* Zhou et al.

2021) that seemed to be present in all the reductions in 2022 and 2023. In 2022 and 2023 it was detected at ~3σ, which is marginal. This source could be just a speckle noise spike in the Hα filter that mimics a faint Hα emission source. Our March 2024 dataset had the highest Strehl of 26% and best contrast to recover CC2 (see Appendix C for contrast curves). However, the 2024 images did not find any evidence of CC2. Therefore, we conclude that the CC2 object is very likely not real, or it simply faded in Hα brightness in 2024. Even a small 2x decrease in Hα would have made it impossible to recover CC2 in 2024, so it might have just faded. Hence, we report on it in this work since it might brighten in the future, but we are very skeptical that it is real. We will need future observations to see if CC2 (sep=0.348" PA=184.0 deg in March 2023) is, in fact, a real protoplanet and not just noise.

*7.4. Search for Inner Planets in the PDS 70 System and its Inner Circumstellar Disk*

Since we did not use one of MagAO-X's Lyot coronagraphs we have a clear view of the core of the PDS 70 system. In fact, our images are not even saturated, with peak counts of only ~22,000 ADU (saturation > 60,000 ADU) in the raw 2s Hα images in 2023 (and similar for 2022 and 2024). So these are excellent datasets to look for new inner planets.

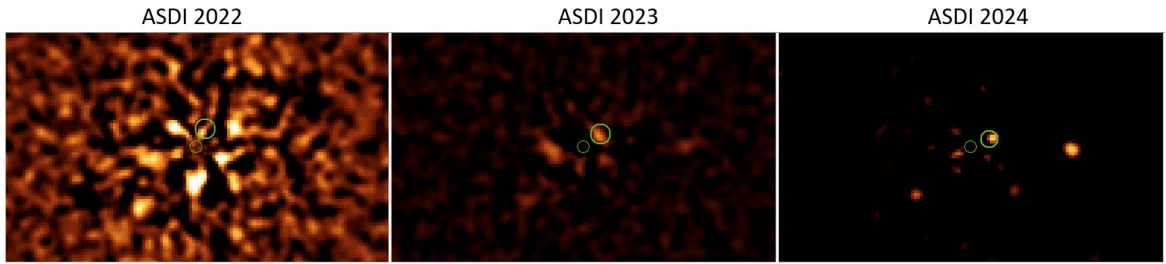

**Fig 14:** the ASDI datasets from Fig. 6 with a deep stretch to show the inner disk area near the star (small light green circle in center). There is only one object inside the orbit of b that could be consistent with a plausible orbit around PDS 70. This r~5.6 au orbit is highlighted by the larger green circle at all epochs. Note that the very bright object "CC3" in the 2023 and 2024 ASDI images track this orbit (large green circle). The object is at 41 mas and PA=305° in 2023 and rotates clockwise (as expected) to 49 mas and PA=295° in 2024. This object is not detected well in the lower quality 2022 dataset. Object CC3 could be a bright clump in the inner disk.

The search for inner planets is complicated by the presence of a bright inner dust disk. The ALMA observations of Francis & van der Marel (2020) find a significant inner circumstellar disk around PDS 70 A. The size of this disk is estimated at 10 au (88 mas semi-major axis).

We only find one new object that might plausibly be consistent with Keplerian co-planar orbital motion around PDS 70 A that also has H$\alpha$ excess. Using the same BKA forward modeling approach of section 5 we find that the object ("CC3") has ASDIcontrast$_{H\alpha}$ of (2.5±0.5)x10$^{-3}$ at 41±5 mas and PA=305±8º in 2023. In 2024, it rotated clockwise (as expected) to 49±6 mas and PA=295±8º with a similar ASDIcontrast$_{H\alpha}$ of (1.5±0.5)x10$^{-3}$. The 9% Strehl of the 2022 dataset was not high enough for a significant detection of CC3 at its predicted ~40 mas separation.

This ~5.6 au orbit is traced by the large green circles in Fig. 14. We find that the brightest "point source" in the 2023 and 2024 ASDI datasets are roughly consistent with a r~5.6 au orbit. That would put this "CC3" object into a 1:8:16 mean motion resonance (a MMR) between CC3:b:c (see Appendix F). An MMR would be the most stable configuration for another planet in this system (see for example Close 2020). However, r~5.6 au is inside the inner disk which is a difficult location for a planet, but perhaps there is a narrow gap in that disk carved out by CC3. The bigger issue with CC3 is that these 41-49 mas separations are close to the peak (~41 mas) of the first Airy ring of the PSF. The peak of the first Airy ring is where AO correction is somewhat unstable and the photon-noise from the bright asymmetric features in that ring can lead to pyKLIP creating "point-sources" from pure noise. For this reason we treat this "CC3" object with a great deal of skepticism. It also has considerable continuum emission and so it might be imbedded as a bright clump in the inner disk. We note that CC3 is also near where one might expect a spiral to be created in the inner disk by b (on the opposite side of star; see theoretical model image in Fig 10).

It is worth noting that *HST* in 2020 also detected a close-in feature "CC1" at separation ~110 mas and PA~310º (Zhou et al. 2021). Moreover, *JWST* in 2023 has recently, tentatively, detected an emission-line object (which Christiaens et al. (2024) call "d?") this object is possibly similar to the r~13.5 au candidate of Mesa et al. (2019), and is found at separation ~115 mas and PA~292º. We do not detect any point sources at any of these wider ~110-115 mas separations (so we cannot confirm that CC1 or "d?" are planets). But all of these objects CC1, "d?", and our "CC3" are all in roughly the same PA~300º "PA zone", and are all very close to the IWA of each instrument. This "PA zone" may have rotated from ~310º in 2020 (Zhou et al. 2021) to PA 292±11º in 2023 (Christiaens et al. 2024), and to ~295±8º in 2024 from this work. Hence, it appears that there is something likely real, quite bright, (with both continuum and Hα emission), and rotating with plausible orbital speeds in the inner disk of PDS 70. It could very well be a bright "spot" or clump in the inner disk, or possibly a real inner giant protoplanet. Future follow up of CC3 will be needed to ascertain if it is indeed real and what its true nature is.

## 8.0 CONCLUSIONS

Sub-mm interferometry (SMA, ALMA etc.) has detected a significant group of large (20-80 au) gaps in many transitional disks (Francis & van der Marel 2020 and references within). We are carrying out a deep survey of the most promising of these wide-gapped disks for Hα emitting protoplanets. We are using the new powerful SDI mode of the extreme AO system MagAO-X for this survey we call MaxProtoPlanetS. We briefly describe how the development of the MagAO-X Hα SDI mode has been, perhaps uniquely well, optimized for the detection of high-contrast Hα protoplanets for 5σ detections at contrasts (ASDIcontrast$_{H\alpha}$) of $1 \times 10^{-3}$ at 50 mas, $7 \times 10^{-4}$ at 100mas, $1 \times 10^{-4}$ at 200 mas, and $2 \times 10^{-5}$ at 300 mas.

Here we present the first Hα protoplanet detections of the MaxProtoPlanetS survey. We recover the PDS 70 b and c protoplanets over a 3 year period (April 2022, March 2023 and March 2024). Due to significant upgrades and better calibration of the AO system our March 2023 data (Strehl 20%; FWHM=27mas; 3.6 hours; seeing 0.45"-0.55") is superior to the April 2022 data (Strehl 9.2%; FWHM=35 mas; 1.4 hours; seeing 0.4"-0.5") even though the seeing was ~20% worse. The addition in 2024 of a new 1024 actuator NCP DM which replaced our previous 97 actuator NCP DM allowed FDPR phase diversity to eliminate most of the NCP aberrations. Hence, in March 2024 we achieved Strehls of 26% and 25 mas FWHM continuously over 2 hours in 0.5" seeing. This 2024 dataset is the sharpest, highest contrast, dataset ever taken of Hα protoplanets.

Our sharp (25-27 mas FWHM) deep (2-3.5hr) March 2023 and 2024 images suggest that there is compact (r~30 mas; r~3 au deprojected) circumplanetary dust surrounding both planets b and c. This dust is the source of compact scattered light at 668 nm in our simultaneously obtained continuum filter images. The detection of compact dust CPDs around protoplanets is an exciting discovery that would benefit from continued observations.

Once we have subtracted this contaminating continuum from the Hα filter (utilizing our custom pyKLIP based ASDI pipeline) we find the Hα line flux of b fell by $(8.1 \pm 1.6) \times 10^{-16}$ ergs/s/cm$^2$ a 4.6x drop in flux from 2022 to 2023 and it continued to be faint in 2024 with just a slight 1.6x rise to an Hα line flux of $(3.64\pm0.87)\times10^{-16}$ ergs/s/cm$^2$ in March 2024. We see evidence that planet c's Hα line flux is also changing, albeit on a smaller scale than for b. We measure a slight (but significant) increase of $(2.74\pm0.51)\times10^{-16}$ ergs/s/cm$^2$ from 2023 to 2024 which is a 2.3x increase in flux. We observe that planet c in 2024 is brighter than b for the first time. Both planets can be significantly variable on ~1 yr timescales, whereas variability on timescales <1 yr were not observed by *HST* (Zhou et al. 2021). This work is the first clear evidence of significant variability of Hα flux from any accreting protoplanets.

We also detect one particularly bright "CC3" Hα excess point source from the inner disk (average separation ~45 mas; at average PA~300º) with orbital motion roughly consistent with a ~5.6 au orbit around PDS 70 A from 2023 to 2024. It is possibly just a PSF artefact. It is also possible "CC3" is a bright clump in the inner disk and the true source of the emission PDS 70 "d?" object detected by *JWST* and the "CC1" object detected by *HST* all of which have PA's (~300º) similar to CC3. Follow-up observations will be required to understand CC3's true nature.

ACKNOWLEDGMENTS

We would like to thank the anonymous referee, whose careful reading of the manuscript and excellent suggestions led to a much improved final manuscript. Laird Close and the MaxProtoPlanetS survey was partially supported by NASA eXoplanet Research Program (XRP) grant 80NSSC18K0441 and is now supported by 80NSSC21K0397 which currently funds the MaxProtoPlanetS survey. Support for this work for Sebastiaan Haffert was provided by NASA through the NASA Hubble Fellowship grant #HST-HF2-51436.001-A awarded by the Space Telescope Science Institute, which is operated by the Association of Universities for Research in Astronomy Inc. (AURA), under NASA contract NAS5-26555. Maggie Kautz and Jialin Wu are supported by NSF Graduate Research Fellowships. Alex Hedglen received a University of Arizona Graduate and Professional Student Council Research and Project Grant in February 2020. Alex Hedglen was partially supported by an Arizona TRIF/University of Arizona "student link" award. We are very grateful for support from the NSF MRI Award #1625441 (for MagAO-X development). The MagAO-X Phase II upgrade program is made possible by the generous support of the Heising-Simons Foundation. The development of pyKLIP is led by Jason Wang and collaborators see https://pyklip.readthedocs.io/en/latest/

*Facilities*: Magellan:Clay (MagAO); Magellan:Clay (MagAO-X)

Corresponding author lclose@arizona.edu

[1,2] Center for Astronomical Adaptive Optics, Department of Astronomy, University of Arizona, 933 N. Cherry Ave. Tucson, AZ 85718, USA

[3] Center for Computational Astrophysics, Flatiron Institute, 162 5th Avenue, New York, New York, USA

[4] Northrop Grumman in Rolling Meadows, Illinois, USA

[5] The Earth and Planets Laboratory (EPL), Carnegie Institution for Science, USA

[6] Amherst College Department of Physics and Astronomy, Science Center, 25 East Drive, Amherst, MA, USA

[7] University of Montreal, Montreal Quebec, Canada

[8] Subaru Telescope, National Observatory of Japan, NINS, 650 N. A'ohoku Place, Hilo, Hawai'I, USA

[9] Wyant College of Optical Sciences, The University of Arizona, 1630 E University Boulevard, Tucson, Arizona, USA

[10] Astrobiology Center, National Institutes of Natural Sciences, 2-21-1 Osawa, Mitaka, Tokyo, Japan

[11] Wyant College of Optical Sciences grad school, The University of Arizona, 1630 E University Boulevard, Tucson, Arizona, USA

[12] Draper Laboratory, 555 Technology Square, Cambridge, Massachusetts, USA

[13] University of Arizona, Department of Physics, Tucson, USA

[14] University of Chile: Santiago, Chile

[15] Starfire Optical Range, Kirtland Air Force Base, Albuquerque, New Mexico, USA

**[16] ETH Zurich Institute for Particle Physics and Astrophysics, HIT J 23.7, Wolfgang-Pauli-Strasse 27, 8093 Zurich, Switzerland**

[17] Department of Physics, National Taiwan Normal University, Taipei 116, Taiwan

## APPENDIX A: The Hα MagAO-X PSF and Strehl Ratio Calculations

The most important input parameter for any high-contrast imaging reduction is the long term PSF of the observation. However, there is almost never a disclosure of the PSF quality that is used in published work, this is particularly true at Hα where the AO PSF can be very low Strehl. Below (Fig. A1) we show what our 3.5 hour Strehl=20%; FWHM=27 mas actually PSF looked like in units of ph/min/pix.

Another, surprisingly rare, disclosure of an imaging manuscript is the final Strehl ratio of the observation, this is never done with Hα imaging papers. However, it is very clear that the achieved Strehl is a very important predictor of SNR of high-contrast imaging (see Fig. 6). We outline here the steps taken to ensure an accurate measurement of the Strehl in our Hα image shown above. The first step is that a perfect PSF was calculated for the pupil mask in use (the pupil clean-up mask; called the MagAO-X "bump mask"), and this was convolved with a 1.35 pixel (8.1 mas) Gaussian to account for CCD charge diffusion at Hα to form the reference PSF. To account for missing flux in the wings of the measured PSF (due to a lack of signal in the wings), a 3 component model was fit to the data. This consisted of a smoothed copy of the reference PSF, and two Moffat profiles. This resultant profile was then integrated and compared to the sum of the measured PSF, yielding a 15% correction of all the missing flux past 0.5" radius in the faint Hα PSF. The Strehl was then estimated by comparing the peak value normalized by the cumulative sum of the measured PSF to the reference PSF, doing so vs. increasing aperture radius. The minimum value along the curve of growth was adopted as the Strehl estimate of 20% for the 1-nm Hα images from 2023 and 9% for 2022 and 26% for 2024. See Fig. 6 to compare these PSFs.

In 2023 our raw 6631 2s Hα images were very consistent in Strehl. We selected 95% of these (6573) that had individual Strehl values (estimated from peak PSF counts) between 15-25% these 6573 were the images selected for the final 3.5 Hα hour exposure PSF shown below (Fig A1). Also the pinned speckle pattern was very stable as well in the 2023 dataset.

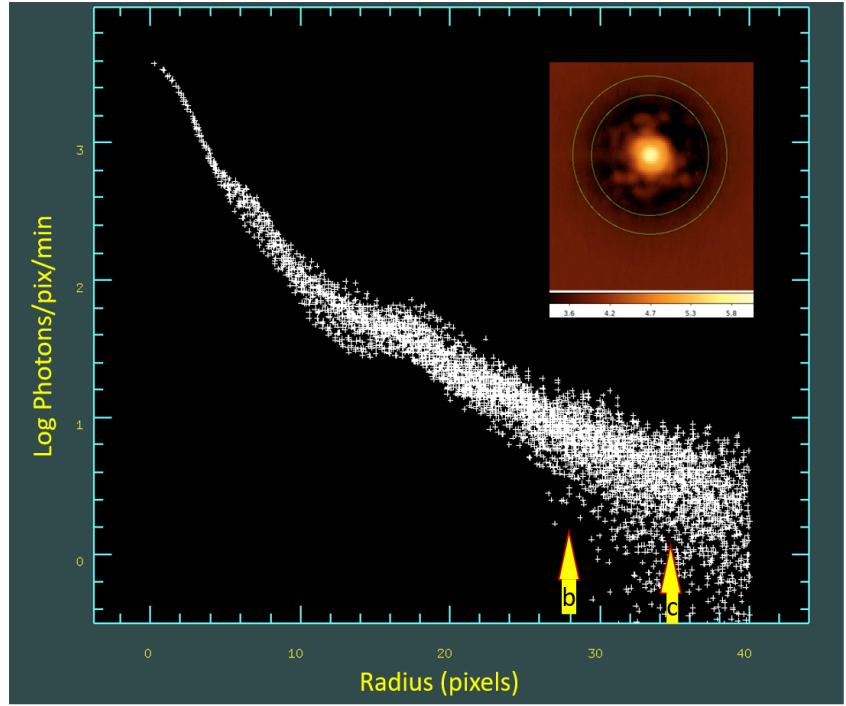

**Figure A1:** Here we see the 3.5 hour PSF of PDS 70 A at Hα of the March 2023 dataset. The units are the log10 of the photons/min/pixel on a median stack of 219x 60s images, the x-axis is in 0.0059" pixels. There is no smoothing or block averaging applied. We show the positions of the peak pixel locations of the b and c planets with the correct contrasts. This image is very helpful in that it shows how pyKLIP is clearly needed to detect the planets because they are ~3x below the noise floor of the PSF. However, we have 219x 60s images and so pyKLIP can easily trace and remove the lower orders in the PSF and through ADI clearly detect both the b and c planets. At the locations of b (155mas) the stellar photons are about equal to those from by the residual read noise, bias and CIC residuals. By the position of c (206 mas) the flux is more dominated by readnoise, bias, and CIC noise residuals which are independent of light from the star as the PSF starts to flatten out. This plot is also very useful as it shows that we expect ~2.6 Hα photon/min/pix from the center planet b pixels and ~1.8 Hα photon/min/pix from planet c (these are very faint signals). **Inset:** we show a log stretch of the PSF with a 18.5pix high pass filter (this leads to a slight dip around the central star, but helps highlight compact speckles) but very little flux is removed from the core of the PSF (or the planet) with a 18.5 pix high pass filter. This slightly negative dip around the PSF core is completely removed by the pipeline during the radial profile subtraction step. The 2 green circles trace the circles that the b and c planets traced out as they moved 137 degrees along these circles during the observations as the sky parallactic angle rotated.

Table A1 is log of all the observations and settings for these PDS 70 observations.

| | 24 April 2022 | 8 March 2023 | 25 March 2024 |
|---|---|---|---|
| | *Environmental* | | |
| Seeing (") | 0.4-0.5" | 0.45-0.55" | 0.4-0.6" |
| Wind (mph) | 13-19; NNE | 7-10; NNE | ~0-2; "N" |
| Photometric sky? | yes | yes | yes |
| | *Adaptive Optics Settings of MagAO-X* | | |
| Number of AO Modes Corrected | 460 | 536 | 624 |
| AO Loop Speed (Hz) | 666 | 1000 | 1000 |
| NCP DM | Alpao DM95 | Alpao DM97 | BMC 1024 (1K) |
| NCP aberration Correction | by eye | LOWFS | FDPR |
| | *Science Camera Features* | | |
| Camera 1 filter : $\lambda_1, \Delta\lambda_1$ (CONT) | 668.0, 8.0 nm | 668.0, 8.0 nm | 668.0, 8.0 nm |
| Camera 2 filter : $\lambda_2, \Delta\lambda_2$ (H$\alpha$) | 656.3, 7.9 nm | 656.3, 1.045 nm | 656.3, 1.045 nm |
| Bump Mask in pupil? | yes | yes | no, open |
| $EM_1$ (CONT) as set on camera 1 | 100 | 100 | 200 |
| $EM_2$ (H$\alpha$) as set on camera 2 | 100 | 300 | 600 |
| EMgain$_{CONT}$ (ADU/e-) | 24.22±0.14 | 24.22±0.14 | 45.84±0.47 |
| EMgain$_{H\alpha}$ (ADU/e-) | 35.46±0.03 | 102.13±0.09 | 196.09±0.17 |
| Readnoise$_1$ rms e- (CONT) | 0.92 | 0.92 | 0.48 |
| Readnoise$_2$ rms e- (H$\alpha$) | 0.48 | 0.16 | 0.08 |
| | *Exposure Times and PDS 70 Observational Parameters* | | |
| Exposure time (DIT) | 2s | 2s | 1s |
| Percentage of raw frames kept | 63.3% | 96.7% | 86.6% |
| Number of raw frames kept | 2393 | 6573 | 7124 |
| Exposure time of combined images | 60x 2=120s | 30x 2=60s | 60x 1=60s |
| #of combined images fed to pyKLIP | 39 | 219 | 118 |
| Total deep exposure time (hours) | 1.3 hr. | 3.6 hr. | 2.0 hr. |
| ADI sky rotation (start→stop: Δdeg) | -45→+51: 96° | -68→+69: 137° | -18→+71: 89° |
| High-Pass (HP) filter value (pix) | 5.333 | 5.333 (19.5 in figs 9,13) | 5.333 (19.5 in figs 9,13) |
| StarFlux$_{H\alpha}$/StarFlux$_{CONT}$ | 1.67 | 0.574 | 0.820 |
| QE$_{CONT}$/QE$_{H\alpha}$ | 14.4/14.3=1.01 | 14.5/14.3=1.01 | 16.8/16.6=1.01 |
| r' mag of PDS 70 A from StarFlux$_{CONT}$ measurements | 11.89±0.04 | 11.80±0.04 | 11.65±0.04 |
| FWHM of H$\alpha$ PSF (deep image) | 29.5 mas | 26.0 mas | 23.6 mas |
| Strehl of H$\alpha$ PSF (deep image) | 9% | 20% | 26% |
| ***Beta(β)*** = (StarFlux$_{H\alpha}$ / StarFlux$_{Cont}$)*(EMgain$_{CONT}$/EMgain$_{H\alpha}$)*( QE$_{CONT}$/QE$_{H\alpha}$) | | | |
| Beta (β) | 1.152 | 0.1375 | 0.1936 |
| SDI "Contrast boost" = 1/β | 0.87x | 7.27x | 5.16x |
| ASDIcontrast$_{continuum}$ = ASDIcontrastH$\alpha$ * β = (ASDI planet flux)/(Star continuum flux) | b=(3.2±0.5)x10$^{-4}$ c=(1.0±0.5)x10$^{-4}$ | b=(6.5±0.6)x10$^{-5}$ c=(5.8±0.4)x10$^{-5}$ | b=(9.1±2.1)x10$^{-5}$ c=(1.2±0.1)x10$^{-4}$ |
| | *pyKLIP parameters and SNR* | | |
| pyKLIP Sectors, Annuli, Modes | 4, 10, 10 | 4, 10, 10 | 4, 10, 10 |
| pyKLIP movement | 0 | 0 (5 in figs 9,13) | 0 (5 in figs 9,13) |
| SNR of b in ASDI image | 5.3 | 10.4 | 4.3 |
| SNR of c in ASDI image | 2.2 | 13.1 | 12.3 |

**Table A1:** Log of all the observations, settings, and parameters used for our PDS 70 observations.

APPENDIX B: Noise Propagation in the Hα Line Flux Calculation

In Fig. B1 (top row) we show how fake negative planets enable accurate BKA forward modeled fluxes for b and c to **be measured**. However, there can be residual speckle/photon noise biases at the positions of the planets, so we need to estimate what the typical speckle noise is at the radii of the planets. In figure B1 (bottom row), we see that (due to additive residual speckle/photon noise in the images) the ring of fake planets all have slightly different final fluxes (despite all have exactly the same initial flux of b or c). We then use standard DAOphot aperture photometry ($r_{app}$=FWHM; sky annuli start=2 FWHM; sky annuli width =3pix) on each fake planet and divide by that of the real planet. We find that the sample of fake planet fluxes normalized to the planet: mean±[sum(flux-mean)$^2$/(N-1)]$^{0.5}$ = 1.11±0.23 for b and 0.978±0.081 for c. So we adopt an error on the ASDIcontrast$_{Hα}$ of b of 23% and 8.1% for c for the 2024 epoch. An identical analysis was carried out for the 2022 and 2023 epochs (ASDIcontrast$_{Hα}$ errors were 14.3% and 44% for b and c in 2022; and 9.5% and 7.5% for b and c in 2023; respectively). The ASDIcontrast$_{Hα}$ values and errors determined this way for all epochs are reported on line 7 of tables 1-3. This observational error term dominates the Hα line flux uncertainty calculated on line 8 of tables 1-3.

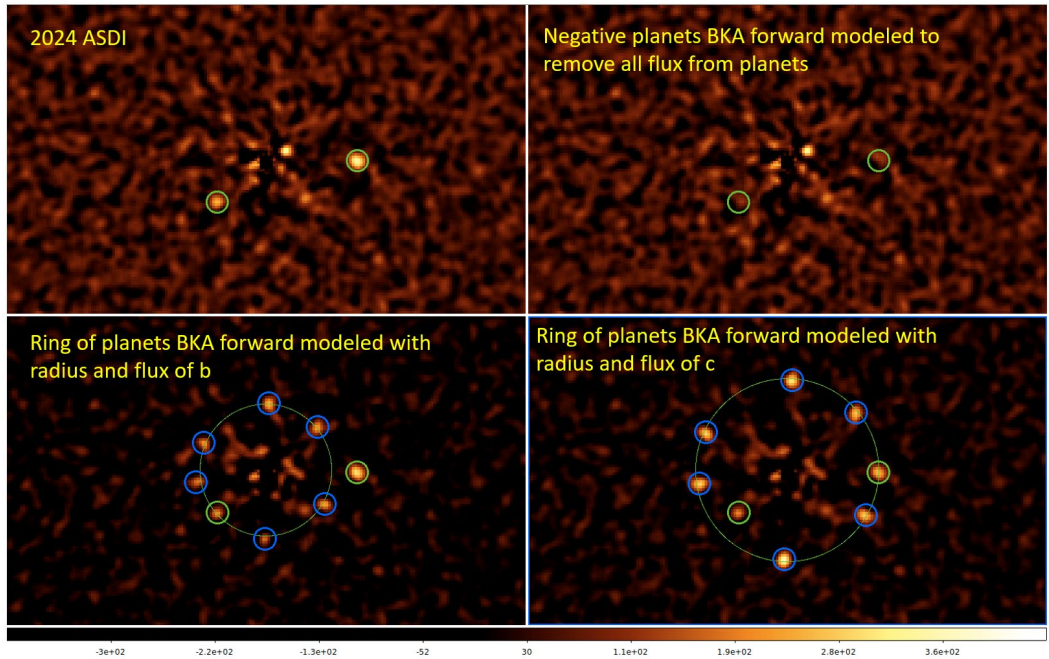

**Figure B1**: The upper images show the original 2024 ASD image. Then to the right is the same dataset after BKA forward modeling of negative fake planets added on top of the real planets. When the astrometry and photometry of the fake planets matches the flux at the positions of both planets is zero. This verifies that the fake planets are correct in flux allowing the ASDIcontrast$_{Hα}$ to be directly measured. On the bottom row we inserted a "ring of fake planets" all at the ASDIcontrast$_{Hα}$ of b (left) and c (right) to estimate the residual noise in the final ASDI images at the radii of the planets.

Below (fig. B2) we show how propagating the pyKLIP BKA forward modeled photometric errors of the planets (Tables 1, 2 and 3) and the photometric errors of PDS 70 A (Table A1), and all the EMgain errors (Table A1) as Gaussian distributions yield the following Gaussian distributions for the Hα line flux. All the flux errors in tables 1-3 were fit in this manner to correctly propagate all errors in the line flux calculations.

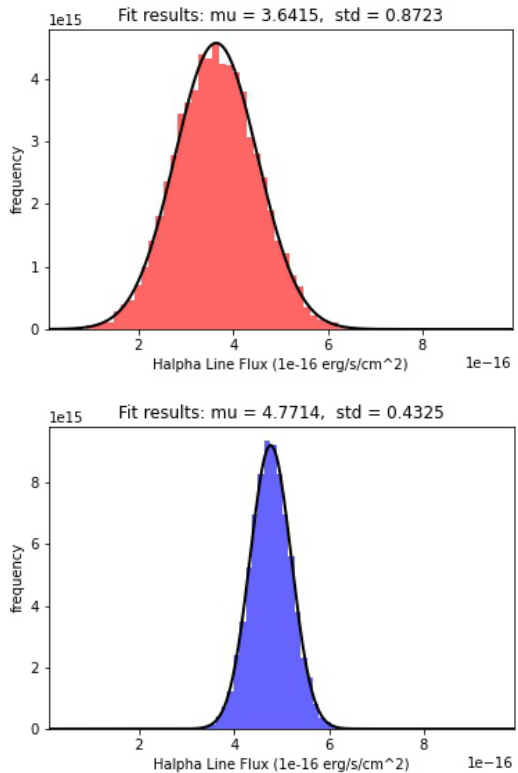

**Fig. B2:** Here we see a full Gaussian propagation of 10,000 random draws of errors with equation 4 of observed uncertainties of the r' flux of PDS 70 A and the those of the measured ASDI contrasts of b and EMgains. It is fit (black Gaussian lines) quite well by a (3.64±0.87)x10$^{-16}$ erg/s/cm$^2$ H$\alpha$ line flux for b (red; top) and by (4.77±0.43)x10$^{-16}$ erg/s/cm$^2$ for c (blue; bottom). The c-b flux difference is (1.13±0.97)x10$^{-16}$ erg/s/cm$^2$, hence c is ~1.2$\sigma$ brighter than b in 2024. This implies that there is ~12% chance that b is actually brighter than c (despite how clearly c looks brighter than b in our 2024 data). Regardless of possibility that these errors are maybe slight overestimates, we adopt these errors throughout this study.

## APPENDIX C: PyKLIP Contrast Curves

Below we present the contrast curves has determined by the pyKLIP package (Wang et al. 2015) for the 5$\sigma$ noise level at the separations plotted on the x-axis (from separations of 0.05" to 0.59"; fully corrected with the weaker significance at small separations of Mawet et al. 2014). The first curve to left in C1 is from the 2022 dataset and then 2023, and 2024 datasets. All of these curves have been rigorously tested for accuracy with fully forward modeled fake planet injections (at known contrasts) into all the raw data and then recovered and the SNR measured to confirm the curves below.

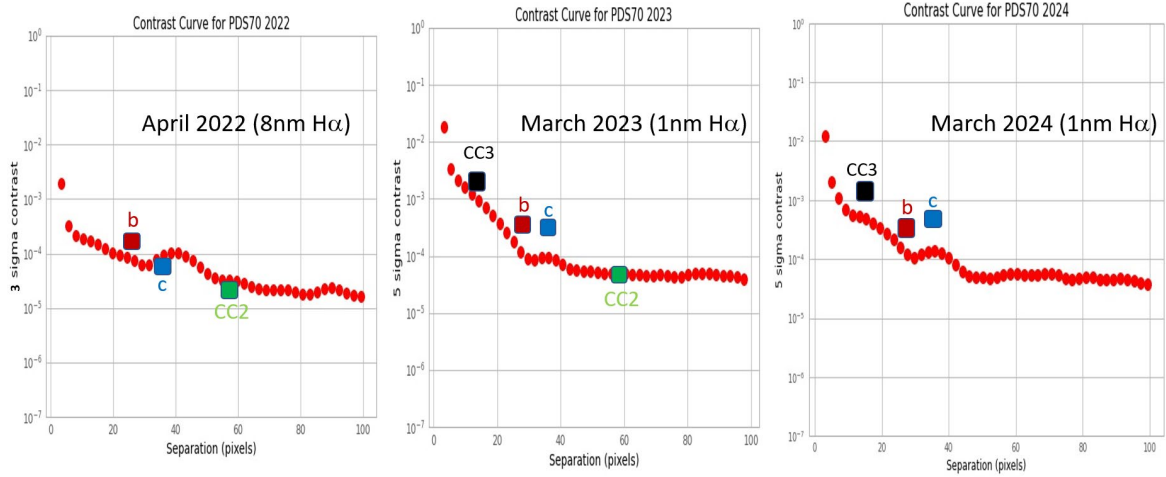

**Fig C1:** ASDI contrast curves (ASDIcontrast$_{H\alpha}$ vs. Separation) from 2022 (3σ) and 5σ for 2023 and 2024 (1pix = 0.0059"). **Left:** we show the 3σ contrast curve for the 2023 dataset with a movement of 0 with 10 KLIP modes after pyKLIP reduction of our 39x 120s images. We also plot the actual detections of planets b (contrast of 2.8x10$^{-4}$; and c (0.9x10$^{-4}$) which were detected at SNR=5.3 and SNR=2.2; respectively. See the right hand panel image in Fig. 2 to see the image for which this curve was generated. **Middle:** the 5σ ASDI 2023 data reduction of 219x 60s images with pyKLIP (movement 0; 10 modes; see right panel of Fig. 3). We also show the detections of planets b (contrast of 4.75x10$^{-4}$) and c (contrast of 4.25x10$^{-4}$) detected at SNR=10.4 and SNR=13.1; respectively. Note, that even though the ASDI Hα flux of b was 4.6x fainter in 2023, the contrasts are larger (easier to detect) than in 2022 because we switched from the 8 nm wide Hα filter in 2022 to the much narrower 1nm filter in 2023. So even if raw ASDIcontrast$_{H\alpha}$ were only marginally increased in 2023 our sensitivity to lower Hα line fluxes was increased by ~8x by our "contrast boost" by using the narrower 1nm filter and better Strehl. **Right:** Here we show the 2024 dataset with b at ASDIcontrast$_{H\alpha}$ = **4.7x10$^{-4}$** at SNR=4.3 and c with ASDIcontrast$_{H\alpha}$ = 6.2x10$^{-4}$ and SNR=12.3, this is based on the image in Fig. 4.

APPENDIX D: Mass Accretion Rate Estimates

We can use the flux values from Table 4 and the equations in section 6.2 to calculate the mass accretion rate ( $\dot{M}_p$ ) for each planet as a function of time. In figure D1 we illustrate these distributions of $\dot{M}_p$. There is a great deal of uncertainty in calculating $\dot{M}_p$ due to uncertainty in the exact form of the power law we should be using in equation 6. Moreover, we only can bound the extinction to $A_p+A_R$~0-3 mag (Zhou et al. 2021). So we also plot upper limits to $\dot{M}_p$ assuming that $A_R$=3 mag and so the flux is suppressed by ~16x. Hence, in the $A_p+A_R$=3 mag case, the true $\dot{M}_p$ is much greater (upper dotted curves) than the $A_p+A_R$=0 mag case (solid lines).

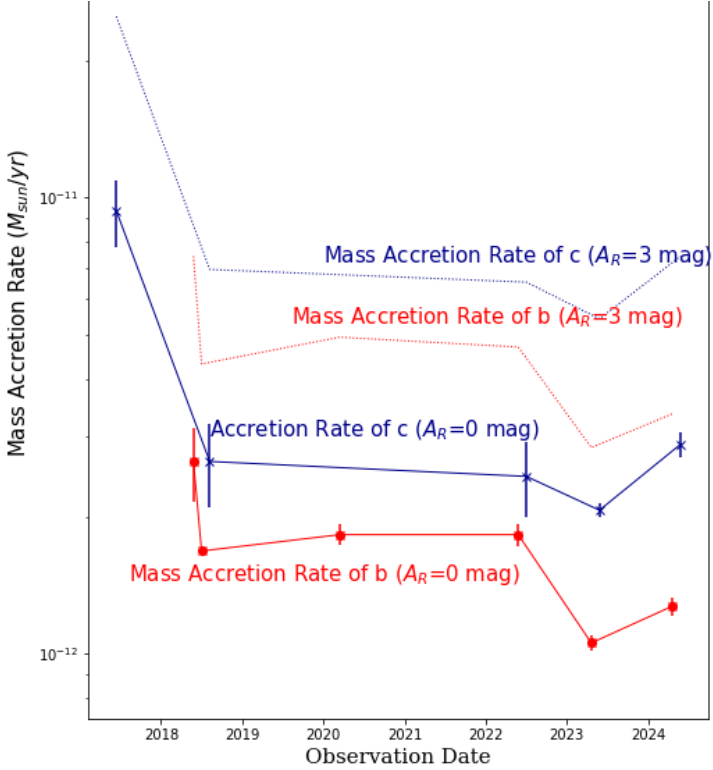

**Fig D1**: The purpose of this plot is to "bound" $\dot{M}_p$ and give a rough order of magnitude of the mass accretion rate of both these protoplanets. We also note from section 6.2 that these values assume $M_p$ b = 4 $M_{Jup}$ and $M_p$ c = 2 $M_{Jup}$ and $R_p$ = 1.3 $R_{Jup}$ for both b and c. There is also ~2x uncertainty in these adopted planet parameters, but as is clear from equation 7, $\dot{M}_p$ varies linearly with $R_p/M_p$ and so the values in Fig. D1 could be off by ~1-4x due to errors in our planetary mass and radius estimates, hence all we can say for certain is that the order of magnitude is $\dot{M}_p$~$10^{-13}$ $M_{sun}$/yr values for both planets where any of the values between the solid and dotted lines in Fig. D1 are possible. *NOTE: these values are corrected for a 24x scaling error in the [published AJ](#) version of this paper. The published version in AJ has some mass accretion rates too low by 24x. This version of figure D1 is correct and all the values in section 6.2 are also correct in this preprint.*

Appendix E: Cartoon of the PDS 70 dust and Hα distribution

In fig E1 we show a cartoon of the dust in the PDS70 system based on the continuum image.

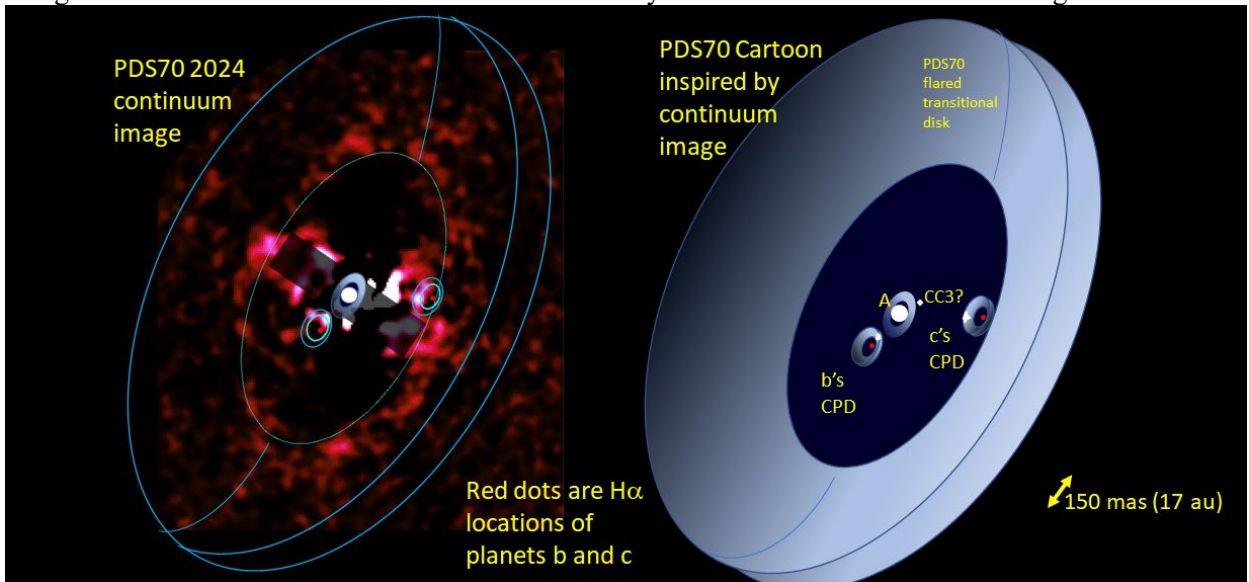

**Figure E1:** A simple cartoon of the dust disks in the PDS 70 system. *Left:* continuum image from 2024 (same data as lower left of Fig. 9) fit to 2 CPDs around each planet (red circles). *Right:* 2 CPDs (radius~1 Hill sphere) are shown with bright starlight from A scattering off the **front edge/lip of each** CPD as observed.

Appendix F: The Predicted Planetary Parameters from the Close (2020) MAG Model of Gap Planets including PDS 70 b, c and d (*see row 7*) reproduced here for reference

| | Orbital semi-major axis (au) | | | Average Projected* separation on-sky (") | | | Planet/Star contrast at ΔmagHα (mag)** | | | Predicted Mass of planet ($M_{jup}$)** | | |
|---|---|---|---|---|---|---|---|---|---|---|---|---|
| Name | $a_1$ | $a_2$ | $a_3$ | $Sep_1$ | $Sep_2$ | $Sep_3$ | $\Delta H\alpha_1$ | $\Delta H\alpha_2$ | $\Delta H\alpha_3$ | $Mp_1$ | $Mp_2$ | $Mp_3$ |
| HD 100453 | 8.97 | 14.23 | 22.59 | 0.08 | 0.13 | 0.21 | 11.11 | **11.61** | 12.12 | 5.88 | 2.94 | 1.47 |
| HD 100546 | 8.67 | 13.76 | 21.84 | 0.07 | 0.11 | 0.17 | 7.42 | 7.92 | 8.43 | 8.52 | 4.26 | 2.13 |
| HD135344B | 15.54 | 24.67 | 39.16 | 0.11 | 0.18 | 0.29 | 9.03 | 9.54 | 10.04 | 6.04 | 3.02 | 1.51 |
| HD 169142 | 7.77 | 12.33 | 19.58 | 0.07 | 0.11 | 0.17 | 7.86 | 8.37 | 8.87 | 6.6 | 3.3 | 1.65 |
| LkCa 15 | 21.52 | 34.16 | 54.22 | 0.12 | 0.19 | 0.30 | 5.50 | 6.00 | 6.51 | 5.28 | 2.64 | 1.32 |
| MWC 758 | 18.53 | 29.41 | 46.69 | 0.11 | 0.18 | 0.29 | 8.87 | 9.37 | 9.88 | 7.08 | 3.54 | 1.77 |
| *PDS 70* | *22.71* | *36.05* | *57.23* | *0.18* | *0.28* | *0.45* | *7.25* | *8.44* | *8.94* | *3.20* | *1.6* | *0.80* |
| UX Tau A | 9.86 | 15.65 | 24.85 | 0.06 | 0.10 | 0.16 | 6.98 | 7.49 | 7.99 | 5.60 | 2.8 | 1.40 |
| V1247 Ori | 19.13 | 30.36 | 48.19 | 0.04 | 0.07 | 0.10 | 7.54 | 8.05 | 8.55 | 7.28 | 3.64 | 1.82 |
| AA Tau | 13.15 | 20.87 | 33.13 | 0.08 | 0.12 | 0.2 | 7.72 | 8.22 | 8.73 | 2.72 | 1.36 | 0.68 |
| AB Aur | 46.62 | 74.00 | 117.47 | 0.26 | 0.41 | 0.66 | 8.68 | 9.18 | 9.69 | 10.2 | 5.12 | 2.56 |
| CQ Tau | 14.94 | 23.72 | 37.65 | 0.09 | 0.14 | 0.22 | 11.96 | **13.48** | 13.98 | 6.52 | 3.26 | 1.63 |
| CS Cha | 11.06 | 17.55 | 27.86 | 0.06 | 0.1 | 0.15 | 6.15 | 6.66 | 7.16 | 5.60 | 2.8 | 1.40 |
| DM Tau[+] | 7.47 | 11.86 | 18.82 | 0.05 | 0.07 | 0.12 | 7.36[+] | 7.86[+] | 8.36[+] | 1.56 | 0.78 | 0.39 |
| DoAr 44[+] | 11.95 | 18.98 | 30.12 | 0.08 | 0.13 | 0.20 | 8.58[+] | 9.09[+] | 9.59[+] | 5.60 | 2.80 | 1.40 |
| GM Aur | 11.95 | 18.98 | 30.12 | 0.06 | 0.10 | 0.15 | 4.88 | 5.39 | 5.89 | 4.04 | 2.02 | 1.01 |
| HD 34282 | 26.00 | 41.27 | 65.51 | 0.06 | 0.10 | 0.16 | 11.21 | 11.72 | **12.09** | 8.44 | 4.22 | 2.11 |
| HD 97048 | 18.83 | 29.89 | 47.44 | 0.08 | 0.13 | 0.21 | 12.28 | **12.68** | 13.04 | 8.68 | 4.34 | 2.17 |
| HP Cha[+] | 14.94 | 23.72 | 37.65 | 0.09 | 0.14 | 0.21 | 8.79[+] | 9.30[+] | 9.80[+] | 3.8 | 1.9 | 0.95 |
| IP Tau[+] | 7.47 | 11.86 | 18.82 | 0.05 | 0.07 | 0.12 | 7.07[+] | 7.57[+] | 8.08[+] | 2.16 | 1.08 | 0.54 |
| RY Lup | 20.62 | 32.73 | 51.96 | 0.08 | 0.13 | 0.21 | 5.61 | 6.11 | 6.62 | 5.6 | 2.8 | 1.40 |
| RY Tau | 8.07 | 12.81 | 20.33 | 0.03 | 0.05 | 0.07 | 6.46 | 6.96 | 7.47 | 9.00 | 4.50 | 2.25 |
| T Cha[+] | 10.16 | 16.13 | 25.6 | 0.06 | 0.1 | 0.15 | 6.65[+] | 7.16[+] | 7.66[+] | 4.48 | 2.24 | 1.12 |

*We note that this is simply an average position, the true position on the sky depends on the unknown orbital phase and so these *sep* values can underestimate the true *sep* by $(a/\pi D)(\pi-2)(1-\cos(inclination))$ and overestimate by $(a/D)[\cos(inclination)-(1+((2-\pi)/\pi)(1-\cos(inclination)))]$ arcsec.

** Assuming $Mp_1=2Mp_2$ and $Mp_2=2Mp_3$. The $\Delta H\alpha_1$ contrasts could have errors of up to 1.0 mag to -0.6 mag and $\Delta H\alpha_2$ contrasts could have errors of +0.5 mag to -0.3 mag if the mass ratios vary from 1.4x to 3x instead of 2x.

Values in **Bold** text are weak accretors and have ΔmagHα calculated by equation 6 (all others use equation 5).

[+] faint $R_A$ >12 mag AO targets have had their contrasts increased by +2mag so they can be compared to the AO sensitivity limits in Fig 8 in Close 2020. If they were observed from space -2 mag should be applied to contrast.